\documentclass[letter,11pt]{article}
\usepackage{jheppub}
\usepackage{tikz}
\usetikzlibrary{decorations.pathmorphing}
\usetikzlibrary{decorations.markings}
\usepackage{xspace}
\usepackage{amsmath}
\usepackage{amssymb}
\usepackage{mathtools}
\usepackage[utf8]{inputenc}
\usepackage{physics}
\usepackage{slashed}
\usepackage{hyperref}
\usepackage[capitalize]{cleveref}

\usepackage{stackengine}

\title{IRC-safe jet flavour at leading power}

\author{Terry Generet}

\affiliation{Cavendish Laboratory, University of Cambridge, Cambridge CB3 0HE, United Kingdom}

\emailAdd{generet@hep.phy.cam.ac.uk}

\preprint{Cavendish-HEP-25/06}

\abstract{We derive the leading power quark mass effects in cross sections involving flavour modulo-2 jets at next-to-next-to-leading order (NNLO) in QCD. Including these leading power terms recovers, up to power corrections, the infrared-collinear-safe massive-quark cross section from the infrared-collinear-unsafe massless-quark one. The method is applicable to all common jet algorithms and significantly more practical than computing fully massive cross sections. We explicitly demonstrate the approach for flavoured jets produced in lepton collisions, inclusive $b$-jet production at the LHC and the associated production of a $b$-jet and a $Z$-boson at the LHC. Through NNLO, we do not observe any breakdown of perturbative convergence resulting from the presence of logarithms of the quark mass, though such effects might become significant at higher orders. The most important feature of our approach is that it does not require any changes to the definition of the jet or its flavour, nor does it modify the definition of the cross section. Consequently, the predictions can be compared to measurements performed using standard jet clustering algorithms, provided that jet flavour is assigned according to the flavour modulo-2 scheme or an unfolding to this scheme is performed, without the need for experimental collaborations to adapt their analyses to some new, infrared-collinear-safe definition of jet flavour, as would be the case for most - if not all - solutions presented in the literature thus far. We further demonstrate that power corrections in the quark mass, which are typically neglected in the literature, can be significant.}

\begin{document} 
\maketitle
\flushbottom

\section{Introduction}\label{sec:intro}

It has long been known \cite{Banfi:2006hf} that none of the standard definitions of jet flavour are well-defined through next-to-next-to-leading order (NNLO) in QCD. Usually, the jets at hadron colliders are defined using the anti-$k_T$ \cite{Cacciari:2008gp}, $k_T$ \cite{Catani:1993hr,Ellis:1993tq} or Cambridge-Aachen \cite{Dokshitzer:1997in,Wobisch:1998wt} jet clustering algorithms. The flavour of the jet is then assigned according to the number of quarks of a specific flavour contained in a given jet. For ease of discussion, but without loss of generality, we will assume throughout this work that bottom is the flavour of interest.

\begin{figure}[t]
\centering
\includegraphics[width=0.99\textwidth]{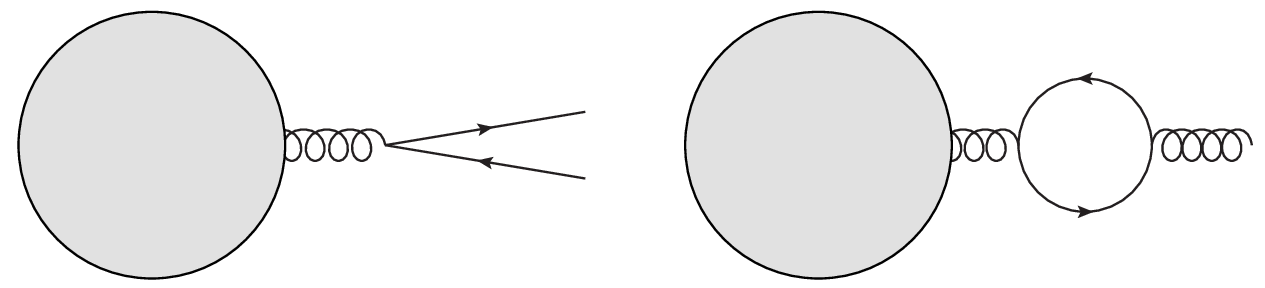}
\caption{Left: the emission of a collinear $b\overline{b}$-pair. Right: the emission of a gluon with a $b$-quark loop. The grey blobs represent some arbitrary process.}
\label{fig:NLOGluonColl}
\end{figure}

Broadly speaking, there are two schemes: in the first scheme, a jet is considered a $b$-jet if it contains at least one $b$-quark or anti-$b$-quark. From now on, we shall call this scheme the `any-flavour' scheme, as was done in ref.~\cite{Caola:2023wpj}. This definition is useful for measurements\footnote{Of course, experimental collaborations do not define flavour in terms of unobservable quarks. Instead, they look for secondary vertices indicative of the presence of $B$-hadrons or study other properties of the jet which correlate with the presence of $b$-quarks (see also ref.~\cite{Behring:2025ilo} for a brief overview). However, this is essentially equivalent to the parton-level scheme described here.}, since it is often difficult to experimentally distinguish jets containing two $b$-quarks from jets containing just one, while distinguishing either from jets containing no $b$-quarks at all is not. However, this definition is rarely used in higher-order calculations, because it is not infrared-collinear (IRC) safe. We say a jet definition is IRC safe if its use does not lead to the non-cancellation of divergences related to soft and/or collinear massless QCD partons. The any-flavour scheme is not IRC safe, because it does not remove all collinear divergences. Indeed, if the final state contains a $b\overline{b}$-pair and this pair becomes collinear (see left panel of fig.~\ref{fig:NLOGluonColl}), the quarks will be clustered into the same jet, which will then be classified as flavoured, since it contains at least one flavoured quark. However, this collinear limit is singular, and thus the cross section diverges (or, if dimensional regularisation \cite{tHooft:1972tcz} with $d=4-2\varepsilon$ is used, it contains uncancelled poles in $\varepsilon$). Note that this divergence is absent if the $b$-quark is taken to be massive. Indeed, all divergences discussed here are only present for massless quarks. It is the choice to treat the quarks as massless which leads to issues of IRC unsafety in jet flavour definitions. Reintroducing the quark mass at leading power renders the cross section finite, which is the underlying motivation for this work.

The second scheme assigns flavour based on whether a jet contains an even or odd number of (anti-)$b$-quarks: a jet with an even number of (anti-)$b$-quarks is considered unflavoured, while a jet containing an odd number of (anti-)$b$-quarks is considered flavoured. Again following ref.~\cite{Caola:2023wpj}, we will refer to this as the `flavour modulo-2' scheme. Using this definition, the poles in $\varepsilon$ cancel: the collinear limit mentioned before still generates a pole, but now the resulting jet is unflavoured and the pole cancels against the pole generated by the contribution shown in the right panel of fig.~\ref{fig:NLOGluonColl}, which likewise generates an unflavoured jet. An analysis of all other singular limits reveals that this definition of jet flavour is IRC safe through next-to-leading order (NLO). For this reason, this scheme has been commonly used in NLO calculations. Since this scheme cannot easily be used for measurements -- as explained above -- this creates a mismatch between the measured cross sections (which use the any-flavour scheme) and the calculated ones (which use the flavour modulo-2 scheme). This mismatch can be accounted for using unfolding techniques. Similar correction steps for unfolding from detector-level flavoured jets to parton-level flavoured jets (see e.g.~refs.~\cite{CMS:2013wql,CMS:2023aim}), from heavy hadrons to heavy quarks (e.g.~ref.~\cite{CMS:2018dxg}) or from flavour-tagged jets to heavy quarks (e.g.~ref.~\cite{LHCb:2014ydc}) have already been used in analyses of LHC data for over a decade. The specific case of unfolding between jet flavour definitions was proposed and performed by theorists in ref.~\cite{Gauld:2020deh} and later also performed by ATLAS in ref.~\cite{ATLAS:2024tnr}.\footnote{It is worth pointing out that, regardless of how well-practised the community is in unfolding, the unfolding between jet flavour definitions is highly sensitive to the modelling of gluon splitting (see e.g.~ref.~\cite{Behring:2025ilo}). Since this modelling as currently implemented in parton showers still leaves something to be desired, it would be ideal to use the any-flavour scheme for theory calculations and eliminate this step. This is a worthwhile avenue for future research, but goes beyond the scope of what this work is trying to achieve.}

\begin{figure}[t]
	\centering
	\includegraphics[width=0.75\textwidth]{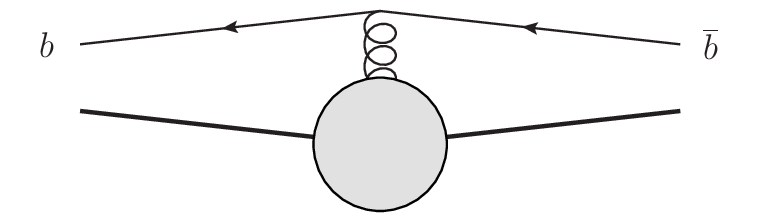}
	\caption{The problematic configuration at NNLO: the two soft $b$-quarks are clustered with different hard partons (thick lines) into two separate jets.}
	\label{fig:NNLOdoublesoft}
\end{figure}

However, starting at NNLO, neither definition of jet flavour is IRC safe when using any of the standard jet clustering algorithms. To illustrate this, we consider the anti-$k_T$ algorithm. However, the analysis for other (IRC-unsafe) algorithms is completely analogous.\footnote{In fact, we will assume the anti-$k_T$ algorithm as the `default' algorithm from now on. In practice, the solution proposed in this paper works for all commonly used algorithms.} Consider the configuration shown in fig.~\ref{fig:NNLOdoublesoft}: a final state containing two hard partons (assumed to be unflavoured) and a soft $b\overline{b}$-pair. Let us assume that the cross section of interest is a two-$b$-jet cross section and that the jets corresponding to the two hard partons satisfy any required cuts. The only way this configuration contributes to the cross section is if the two $b$-quarks are clustered with different hard partons into two separate jets. The angular phase space for such a configuration is of course non-zero, so the integration over the angles of the $b$-quarks will yield a finite (i.e.~non-zero) factor. This leaves the integration over the energies of the $b$-quarks. However, while there is no divergence associated with either quark on its own becoming soft, if both quarks become soft simultaneously, this leads to the well-known double-soft singularity for quark-anti-quark pairs. If this divergence is dimensionally regulated, the corresponding pole will not cancel when summing over all contributions.

This issue is well-known and was first identified in ref.~\cite{Banfi:2006hf}. Since then, and especially in recent years, a number of solutions have been proposed \cite{Banfi:2006hf,Caletti:2022hnc,Caletti:2022glq,Czakon:2022wam,Gauld:2022lem,Caola:2023wpj,Larkoski:2025afg}.\footnote{Recent progress in fragmentation calculations \cite{Czakon:2021ohs,Czakon:2022pyz,Czakon:2024tjr,Generet:2025bqx} has also opened up the possibility to use identified heavy-flavour hadrons instead of flavoured jets to establish the presence of heavy flavour in processes through NNLO.} These all involve modifying the jet (flavour) definition somehow: either the jet definition is modified in a way that guarantees that a soft $b\overline{b}$-pair is clustered into the same jet \cite{Banfi:2006hf,Czakon:2022wam}, in which case the kinematics of the jet(s) will not be identical to what would have been obtained using the anti-$k_T$ algorithm, or the definition of jet flavour is changed \cite{Caletti:2022hnc,Caletti:2022glq,Gauld:2022lem,Caola:2023wpj}, in which case the jet kinematics might not change, but this usually involves some extra step(s). There have even been proposals to modify the definition of the cross section itself \cite{Larkoski:2025afg}. Generally, these solutions introduce some new parameter(s) used to define the jet (on top of the usual jet radius $R$). Needless to say, the introduction of extra free parameters should be avoided. Moreover, if these solutions are to be applied in experiment, then the flavour information entering e.g.~the flavoured jet algorithms would have to be at the level of hadrons or their decay products, since this is what is experimentally accessible. However, theory predictions at the precision frontier typically do not incorporate hadronisation effects, meaning the flavoured jet algorithms applied in the predictions must make use of parton-level flavour information, e.g.~the momenta of $b$-quarks, which is experimentally inaccessible. This mismatch complicates the comparison between theory and experiment.\footnote{Of course, unflavoured jet cross sections are affected by hadronisation as well, and these effects must be corrected for, e.g.~using an unfolding procedure, in order to perform accurate comparisons. The mismatch described here is just an additional effect that must be accounted for.} Finally, there are many solutions, all with their own merits and drawbacks, and it is unclear which should be chosen as the new standard moving forward.\footnote{See also ref.~\cite{Behring:2025ilo} for a study comparing most of these proposals quantitatively.} And once this choice has been made, adapting to this new standard may involve significant effort. In the meantime, the resulting uncertainty has different groups using different algorithms, complicating comparisons of results.

In this work, we propose how this issue can be avoided entirely, allowing the community to keep using the tried and trusted algorithms. We will show that this involves only minor effort from the side of theory collaborations, while experimental collaborations can keep clustering using anti-$k_T$, though an unfolding to the flavour modulo-2 scheme remains necessary. In section \ref{sec:Solution}, we will explain the proposed solution. Some numerical checks will be presented in section \ref{sec:Checks}, followed by two phenomenological studies in section \ref{sec:Pheno}. Finally, we will conclude in section \ref{sec:Conclusion}.

\section{Nature of the problem and its solution}\label{sec:Solution}

As mentioned in the introduction, the problem (at NNLO) boils down to this: there is an uncancelled singularity coming from soft $b\overline{b}$-pairs. Clearly, this singularity is regulated in reality by the non-zero $b$-quark mass. In fact, the only reason why $b$-jet (and $c$-jet) cross sections can be calculated perturbatively is because of the non-zero $b$-quark (and $c$-quark) mass. Strange jets, for example, will get contributions from non-perturbative $g\to s\overline{s}$ splittings, which need to be modelled empirically, using e.g.~a parton shower. Thus, in both cases\footnote{Or all three cases, if one counts top-jets.} where perturbative calculations can be used in the first place, the problem can be avoided if the corresponding quarks are simply given their physical mass. This shows that, in essence, the problem is a consequence of the choice to treat these heavy quarks as massless.

Of course, this is no news to the theory community. In fact, the $pp\to W+2b\mathrm{-jet}$ cross section was computed in ref.~\cite{Buonocore:2022pqq} at NNLO using massive quarks precisely for this reason. However, there are two arguments commonly used to argue why this cannot be the standard. First of all, keeping the $b$-quark mass non-zero introduces another scale to the problem. It is well-known that the more scales are present in a process, the more difficult the computation of the loop amplitudes becomes. The two-loop amplitudes are typically the bottleneck for modern NNLO computations, so making them more complicated to compute would delay the entire NNLO computation.\footnote{To get around this, a massification procedure was used in ref.~\cite{Buonocore:2022pqq} to obtain -- up to power corrections -- the two-loop amplitudes involving massive $b$-quarks from those with massless $b$-quarks.} Additionally, a non-zero mass can also significantly affect the numerical stability and convergence of the calculations. For these reasons, it is clear that just doing the calculation for massive quarks is not a practical solution.

The second reason is that cross sections involving massive quarks contain logarithms of the $b$-quark mass over the hard scale of the process, $\ln(Q^2/m_b^2)$, which can grow arbitrarily large at high energies, spoiling the perturbative convergence of the cross section. In fact, if one considers sufficiently exclusive cross sections, e.g.~distributions at the level of $b$-quarks, then the leading logarithms at order $n$ can behave as poorly as $\alpha_s^n\ln^{2n}(Q^2/m_b^2)$.\footnote{An example of this worst-case behaviour is the energy spectrum of $b$-quarks in Higgs-boson decays \cite{Gaggero:2022hmv}.}

Despite these concerns, our solution to the jet flavour problem is to keep the $b$-quark mass non-zero, at least at leading power (LP). In the following, we will argue how the resulting additional computational complexity can be minimised to be essentially negligible and how the logarithmic behaviour is actually unproblematic for standard jet definitions.

To begin, we recall that the dependence of the cross section on the $b$-quark mass factorises, up to power corrections. This factorisation allows one to obtain the massive result from the massless one using a few process-independent ingredients. Most importantly, this restores the logarithmic dependence of the cross section on the $b$-quark mass and we will be referring to these terms as `logs', though it is understood that the massification also introduces mass-independent terms. To give a common example, there are logs coming from loops associated with the renormalisation of the strong coupling, which can be reintroduced using the $\alpha_s$ decoupling constant \cite{Weinberg:1980wa,Ovrut:1980dg}. There are also logs associated with quasi-collinear configurations in the initial or final state, which can be reintroduced using the corresponding matching conditions for parton distribution functions (PDFs) \cite{Aivazis:1993pi,Buza:1995ie,Buza:1996wv,Bierenbaum:2007qe,Bierenbaum:2008yu,Bierenbaum:2009zt} or fragmentation functions (FFs) \cite{Mele:1990cw,Melnikov:2004bm,Mitov:2004du,Cacciari:2005ry,Neubert:2007je}, respectively. In this way, the leading dependence on the $b$-quark mass can be reintroduced in a process-independent (i.e.~automated) way, without the need to compute any truly massive cross sections. Importantly, these process-independent ingredients (e.g.~the perturbative FFs) also contain poles in $\varepsilon$ which exactly cancel the corresponding poles (e.g.~collinear poles in the final state) in the massless cross section. In general, there is a correspondence between poles in the massless cross section and logs in the massive one. Intuitively, this correspondence arises because the process-independent ingredients only contain (aside from some arbitrary renormalisation scale $\mu$) one scale: $m_b$. Therefore, their entire dependence on $m_b$ is of the form $(\mu^2/m_b^2)^{n\varepsilon}$, which produces logs according to the same pattern as the $\varepsilon$-poles present in the process-independent ingredient, which in turn match the poles of the massless cross section.

Many of these logs are already routinely included and resummed in cross sections. In particular, this includes the logs coming from the $\alpha_s$ decoupling constant and the logs coming from the PDFs. Both are included and resummed in all standard PDF sets available. Therefore, they are already included in massless cross section calculations and do not spoil the perturbative convergence (since they have been resummed). The case of collinear final state logs (i.e.~FFs) is even simpler: they can be omitted entirely.

\begin{figure}[t]
	\centering
	\includegraphics[width=0.99\textwidth]{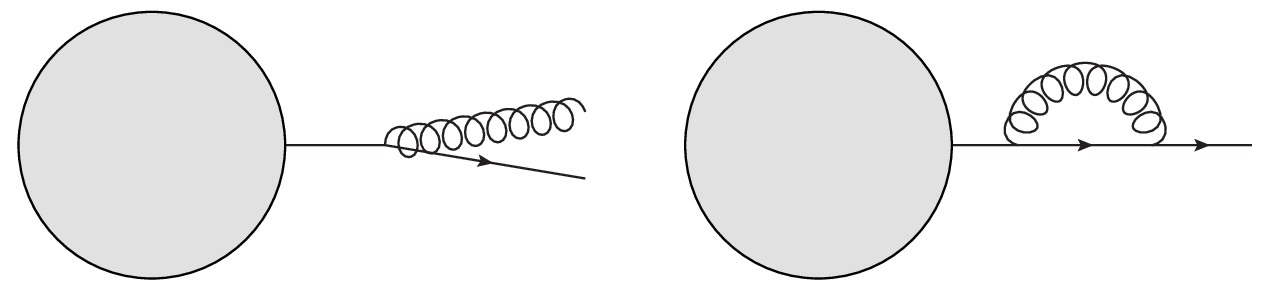}
	\caption{Left: the collinear emission of a $b$-quark and a gluon. Right: the emission of a $b$-quark with a QCD loop. The grey blobs represent some arbitrary process.}
	\label{fig:NLOQuarkColl}
\end{figure}

To see this, consider all possible cases of collinear logs at NLO. The first pair of cases was already shown in fig.~\ref{fig:NLOGluonColl}: a gluon splitting into a collinear $b\overline{b}$-pair and a gluon with a $b\overline{b}$ loop. In the first case, both quarks will be clustered into the same jet, since they are collinear. Since we are using the flavour modulo-2 scheme, they thus do not affect the flavour of any jet. In the second case, the jet flavours are also unaffected, since the particle is a gluon. Thus the logs of these two contributions cancel, provided that they are the same up to a sign. This is of course guaranteed by the Kinoshita-Lee-Nauenberg (KLN) theorem \cite{Kinoshita:1962ur,Lee:1964is} and can be easily checked using the explicit expressions for the one-loop FF matching conditions. The same holds for the other pair of cases, shown in fig.~\ref{fig:NLOQuarkColl}: the logs generated by the emission of a real gluon off a heavy quark exactly cancel those generated by virtual contributions (after integration over the momentum fraction). By the KLN-theorem, this holds to any order: all final-state (soft-)collinear logs cancel as long as one is insensitive to the distribution of collinear momentum among the collinear partons (i.e.~one integrates over the momentum fractions) and to collinear $g\to b\overline{b}$ splittings (i.e.~one uses the flavour modulo-2 scheme).\footnote{While these logs do not cancel for the any-flavour scheme (and thus need to be included), this is not a problem in practice, since the unfolding to the flavour modulo-2 scheme is performed using parton showers \cite{Gauld:2020deh,ATLAS:2024tnr}. These can treat the $b$-quark as massive (avoiding the need to massify anything) and resum the logs at the same time (avoiding the presence of large collinear logs). It is also possible to use a parton shower with massless $b$-quarks (as was done in refs.~\cite{Gauld:2020deh,ATLAS:2024tnr}), since any shower tuned to data should at least effectively capture the mass effects.} Thus, all collinear logs either cancel or are already resummed by default. In particular, this also includes soft-collinear logs.\footnote{Soft-collinear logs related to the initial state (i.e.~soft splittings to $b$-quarks at low $x$) are typically not resummed, but they are phenomenologically irrelevant.} This only leaves soft logs, which first show up at NNLO and thus the cross section only behaves as $\alpha_s^n\ln^{n-1}(Q^2/m_b^2)$ at worst, as will be discussed further below. The logarithmic enhancement of mass effects is thus significantly weaker than the naive expectation.

Since all (soft-)collinear logs have been addressed, this only leaves the soft logs. In the following, we will completely ignore the presence of the other logs, since they have already been addressed. Although this has never been discussed in the literature to the best of our knowledge, the soft logs can also be factorised in analogy to perturbative fragmentation functions. The factorisation looks as follows:
\begin{align}
\mathcal{F}\big(\sigma(m_b)\big) {}={}& \mathcal{F}\big(S_\emptyset(m_b)\otimes\sigma(m_b=0)\big) + \mathcal{F}\big(S_{b\overline{b}}(m_b,p_b,p_{\overline{b}})\otimes\sigma(m_b=0)\big)\notag\\&{}+{} \mathcal{F}\big(S_{b\overline{b}b\overline{b}}(m_b,p_{b1},p_{\overline{b}1},p_{b2},p_{\overline{b}2})\otimes\sigma(m_b=0)\big) + ... + \mathcal{O}(m_b/Q)\;,
\end{align}
where the ellipses are an infinite sequence of analogous contributions with an increasing number of $b\overline{b}$-pairs. $\mathcal{F}$ is a measurement function, which among other things implements the jet (flavour) definition and applies a cut on the number of flavoured jets. The $\otimes$ symbol indicates two things. First of all, it indicates that the particles in the subscript of the function $S$ are inserted into the final state observed by $\mathcal{F}$ with corresponding momenta as shown in the arguments of the function $S$. I.e.: the function $S_\emptyset$ leaves the final state untouched, while $S_{b\overline{b}}(p_b,p_{\overline{b}})$ adds a $b\overline{b}$-pair to the final state with momenta $p_b$ and $p_{\overline{b}}$, etc. Important to note is that these momenta should be treated as infinitesimal, i.e.~they do not affect the kinematics of the other particles and enter the clustering algorithm and cuts as soft partons. This is again in analogy to perturbative fragmentation, where the fragment is taken to be strictly collinear to the fragmenting parton. Second of all, the $\otimes$ symbol indicates that the $S$ functions factorise from the cross section up to colour correlations. This is in full analogy to the way soft singularities of cross sections factorise up to colour correlations. The integration over all momenta is implicit. Since this factorisation assumes that the $b$-quark mass and the momenta introduced by the $S$ functions are small compared to the hard scale of the process, the momenta of the other partons can equivalently be considered to be infinitely large. As such, the integration of the momenta introduced by the $S$ functions should extend to infinity and there is no need to consider the effect of the recoil of the soft quarks on the other partons. The former point will be further explained when the computation of the $S$ functions is discussed below, but is once again fully analogous to the case of perturbative fragmentation, where, despite being `collinear', the relative transverse momenta of the collinear partons are integrated from zero to infinity when computing the perturbative fragmentation functions.

Clearly, the $S$ function for $n$ $b$-quark pairs is only non-zero starting at N$^{2n}$LO. At the same time, a flavour-insensitive cross section is insensitive to mass effects, so
\begin{equation}
	\sum_{i=0}^\infty\int dp_{b1}\int dp_{\overline{b}1}...\int dp_{bi}\int dp_{\overline{b}i}S_{\underbrace{b\overline{b}...b\overline{b}}_{\text{i pairs}}} = \mathbf{1}\;,
\end{equation}
where we suppressed the arguments of the $S$ functions and $\mathbf{1}$ is the identity matrix in colour space, i.e.~it acts trivially on $\sigma$. Through NNLO, this gives
\begin{equation}\label{eq:S0SbbRelation}
	S_\emptyset = \mathbf{1}-\int S_{b\overline{b}} dp_{b}dp_{\overline{b}}+\mathcal{O}(\alpha_s^4)\;.
\end{equation}

\begin{figure}[t]
	\centering
	\includegraphics[width=0.99\textwidth]{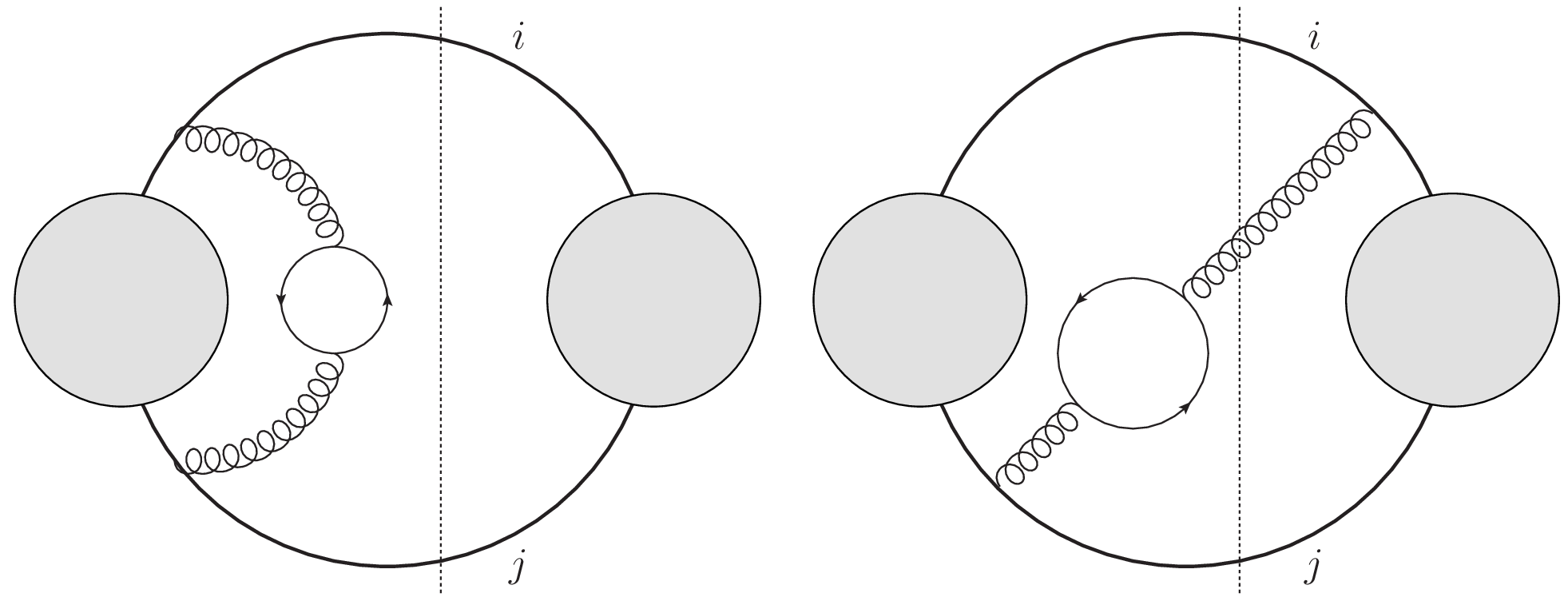}
	\includegraphics[width=0.49\textwidth]{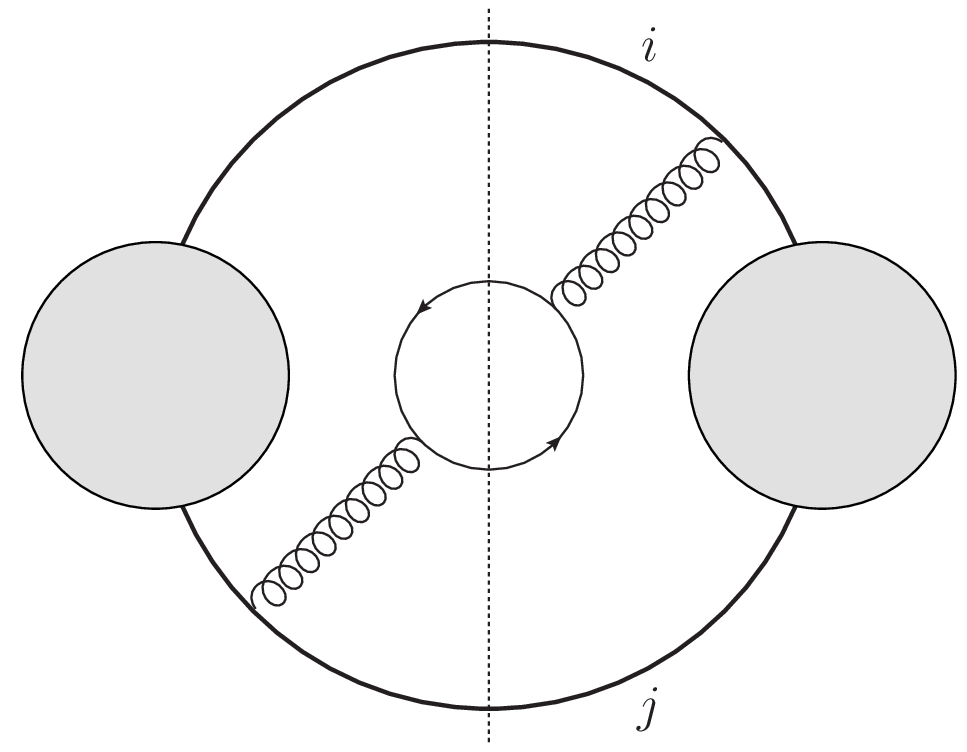}
	\caption{Top-left: a double-virtual diagram which contributes soft mass logs. The thick lines correspond to arbitrary external partons $i$ and $j$, while the dashed line represents the cut. Top-right: the corresponding real-virtual diagram. Bottom: the corresponding double-real diagram. The existence of complex conjugate diagrams is understood for the virtual contributions.}
	\label{fig:NNLOContributions}
\end{figure}

At NNLO, there are three contributions introducing soft logs: one double-virtual, one real-virtual and one double-real, shown in the top-left, top-right and bottom panel of fig.~\ref{fig:NNLOContributions}, respectively. Here, eikonal vertices should be used to attach the gluons to the external partons $i$ and $j$. The double-virtual and real-virtual contributions will modify a cross section without affecting the flavour assignments of the Born jets. These correspond to contributions to $S_\emptyset$. The double-real emission changes the flavour assignments depending on the direction (and possibly the relative magnitude) of the $b$-quark momenta. This is a contribution to $S_{b\overline{b}}$.

The real-virtual contribution is trivial: the $b$-quark bubble reduces to a multiplicative factor. The rest corresponds to the emission of a single gluon. The integral over its energy is scaleless, so the contribution vanishes.

Moving on to the double-virtual contribution, it can in principle be computed analytically using the diagram shown in the top-left panel of fig.~\ref{fig:NNLOContributions}.\footnote{Note that two distinct cases would need to be considered: one where $i\neq j$ and one where $i=j$.} However, we here refrain from performing this exercise. Instead, we use the relation given in eq.~\eqref{eq:S0SbbRelation} to obtain it numerically. I.e.~after deriving the NNLO expression for $S_{b\overline{b}}\otimes\sigma$, we will compute $S_\emptyset\otimes\sigma$ by negation and integration over the soft momenta $p_b$ and $p_{\overline{b}}$, while leaving the final state intact (i.e.~without inserting the soft quarks into the final state). In part, this is to reduce the amount of analytical integration that needs to be performed in order to implement our scheme -- indeed, no analytical integration will be necessary at all -- but there is also a more significant reason which will become apparent when discussing the implementation of $S_{b\overline{b}}\otimes\sigma$ further below.

Moving on to the real contribution, the squared matrix element factorises as follows:
\begin{equation}
\lvert M_{b,\overline{b},a_1...}\rvert^2 \sim (4\pi\alpha_s)^2T_F\sum_{i,j}\mathcal{I}_{ij}(p_b,p_{\overline{b}},m_b)\big\langle M_{a_1,...}\rvert\mathrm{\textbf{T}}_i\cdot \mathrm{\textbf{T}}_j \lvert M_{a_1,...}\rangle\;,
\end{equation}
where we used the same notation for colour correlators as in ref.~\cite{Czakon:2014oma}. This form is of course identical to the case of a massless quark pair. The double-soft function for massive quarks $\mathcal{I}_{ij}(q_1,q_2,m)$ is given by
\begin{equation}
	\mathcal{I}_{ij}(q_1,q_2,m)=\frac{(p_i\cdot q_1)(p_j\cdot q_2)+(p_j\cdot q_1)(p_i\cdot q_2)-(p_i\cdot p_j)[m^2+(q_1\cdot q_2)]}{[m^2+(q_1\cdot q_2)]^2[p_i\cdot(q_1+q_2)][p_j\cdot(q_1+q_2)]}\;.
\end{equation}
Note that for $m=0$, the massless result is trivially recovered. $\mathcal{I}_{ij}(q_1,q_2,m)$ can be derived in full analogy to the massless version by studying the behaviour of amplitudes when both heavy-quark momenta and the quark mass uniformly approach zero, i.e.~$q_1\to\lambda q_1$, $q_2\to\lambda q_2$ and $m\to\lambda m$ with $\lambda\to0$.

At NNLO, the function $S_{b\overline{b}}$ is simply the operator which replaces
\begin{equation}
	\lvert M_{a_1...}\rvert^2 \to (4\pi\alpha_s)^2T_F\sum_{i,j}\mathcal{I}_{ij}(p_b,p_{\overline{b}},m_b)\big\langle M_{a_1,...}\rvert\mathrm{\textbf{T}}_i\cdot \mathrm{\textbf{T}}_j \lvert M_{a_1,...}\rangle\;,
\end{equation}
and adds the two $b$-quarks to the final state. Technically, the $S$ functions should correspond to the difference between the massive and massless cross sections. Therefore $S_{b\overline{b}}$ should correspond to the difference between the double-soft limit of the massive and massless cross sections. In this case, the cross section is proportional to $\mathcal{I}_{ij}(p_b,p_{\overline{b}},m_b)-\mathcal{I}_{ij}(p_b,p_{\overline{b}},0)$. This also highlights that the massification removes the double-soft poles from the massless cross section, since the double-soft limit is explicitly subtracted. This view should be compared to the definition of perturbative fragmentation functions employed in refs.~\cite{Melnikov:2004bm,Mitov:2004du}, which is likewise given by the difference between the massive and massless cases. At sufficiently high energies of the soft $b$-quarks (but by assumption still much smaller than the hard scale of the process), the difference between the massive and massless soft functions gives a vanishing contribution, so what is changed by including $S$ truly is only the soft behaviour. At the same time, the massless contribution vanishes when integrated over the energies, since it is scaleless. It can therefore be omitted, leaving only the contribution proportional to $\mathcal{I}_{ij}(p_b,p_{\overline{b}},m_b)$.

Keeping the massless term for now, the function $S_{b\overline{b}}$ contains a soft pole. In the case of perturbative FFs, the corresponding collinear poles could be treated analytically, since the calculation of the perturbative FFs can be performed analytically: all momenta are integrated over, up to the integration over the longitudinal momentum fraction. In the case of $S_{b\overline{b}}$ at NNLO, the momenta cannot be integrated over analytically, since their directions and relative magnitude\footnote{While the absolute magnitude of the soft momenta is irrelevant (since they are kinematically treated as infinitesimal), their relative magnitude does matter for some jet algorithms, such as the $k_T$ algorithm.} affect the result of the jet clustering. At the same time, the only integration which contains a soft singularity is the integral over the absolute magnitude of the soft momenta. One might e.g.~parametrise the momenta using $x_S$ and $E_S$ like $|\vec{p}_b|=x_S E_S$, $|\vec{p}_{\overline{b}}|=(1-x_S) E_S$, $0\leq x_S\leq1$, $E_S\geq0$, in which case the soft divergence is purely due to the integration over $E_S$. Since the absolute magnitude of the soft momenta is kinematically irrelevant, this integral could in principle be performed analytically, enabling the use of dimensional regularisation to deal with the divergence. However, we instead prefer to perform the integral numerically, using subtraction techniques to deal with the singularity. When omitting the scaleless massless contribution, the soft singularity is traded for a UV singularity at $E_S=\infty$. Studying the integrand reveals the scaling of the cross section in this limit:
\begin{align}\label{eq:UVofSoft}
\frac{d^{d-1}p_b}{E_b}\frac{d^{d-1}p_{\overline{b}}}{E_{\overline{b}}}&\mathcal{I}_{ij}(p_b,p_{\overline{b}},m) = d^{d-2}\Omega_bd^{d-2}\Omega_{\overline{b}}\frac{d\lvert\vec{p}_b\rvert\lvert\vec{p}_b\rvert^{d-2}}{E_b}\frac{d\lvert\vec{p}_{\overline{b}}\rvert\lvert\vec{p}_{\overline{b}}\rvert^{d-2}}{E_{\overline{b}}}\mathcal{I}_{ij}(p_b,p_{\overline{b}},m) \notag\\{}\to{}& d^{d-2}\Omega_bd^{d-2}\Omega_{\overline{b}}\frac{dE_S}{E_S^{1+4\varepsilon}}dx_S\big(x_S(1-x_S)\big)^{1-2\varepsilon}\mathcal{I}_{ij}(x_S\hat{n}_b,(1-x_S)\hat{n}_{\overline{b}},0)\;,
\end{align}
where we performed the variable transformation to $x_S$ and $E_S$ and introduced the versors $\hat{n}_b$ and $\hat{n}_{\overline{b}}$, defined via $p_i=p_i^0\hat{n}_i$ in the massless limit. In the UV, the soft momenta can of course be treated as massless. To perform this integral, we use a variation on the usual trick
\begin{equation}
\int_0^1\frac{f(x)}{x^{1+a\varepsilon}}dx = \int_0^1\frac{(f(x)-f(0))}{x^{1+a\varepsilon}}dx-\frac{1}{a\varepsilon}f(0) = \int_0^1\bigg[-\frac{1}{a\varepsilon}\delta(x)+\frac{1}{x^{1+a\varepsilon}_{{}+{}}}\bigg]f(x)dx\;,
\end{equation}
where we introduced the usual plus-distribution. In our case, the integral extends from $0$ to infinity and the subtraction needs to happen at infinity, not zero. Additionally, the high-energy approximation shown in eq.~\eqref{eq:UVofSoft} contains both a UV and a soft divergence, and so is not suitable as a subtraction term. Instead, we can use the following formula:
\begin{align}\label{eq:UVsubtraction}
\int E_S^{3-4\varepsilon}&\mathcal{I}_{ij}(p_b,p_{\overline{b}},m)dE_S {}={}\\& \int\bigg[E_S^{3-4\varepsilon}\mathcal{I}_{ij}(p_b,p_{\overline{b}},m)-\frac{E_S^A}{(R^B+E_S^B)^{(A+1+4\varepsilon)/B}}\mathcal{I}_{ij}(x_S\hat{n}_b,(1-x_S)\hat{n}_{\overline{b}},0)\bigg]dE_S\notag\\&{}+R^{-4\varepsilon}\frac{\Gamma(\frac{A+1}{B})\Gamma(\frac{4\varepsilon}{B})}{B\:\Gamma(\frac{A+1+4\varepsilon}{B})}\mathcal{I}_{ij}(x_S\hat{n}_b,(1-x_S)\hat{n}_{\overline{b}},0)\;,\notag
\end{align}
where $R$, $A$ and $B$ are free parameters. A correct result does not depend on them, which is a valuable test of any implementation of this scheme. The first term is free of singularities and poles in $\varepsilon$, so it can be integrated numerically in four dimensions. The integrated subtraction term does contain an $\varepsilon$-pole, and so should be evaluated in $d$ dimensions. However, the singular $E_S$ integration has been performed analytically and is no longer an issue. The $d$-dimensional nature of this term means great care must be taken to correctly include all $\varepsilon$-dependent factors coming from the phase space measure. It also means that the integration over the angles of the soft quarks must be performed in $d$ dimensions. This can be done by promoting one of the momenta to five dimensions and the second to six dimensions, as described in ref.~\cite{Czakon:2014oma}.

This last point provides another valuable cross-check of results. The definition of normal four-dimensional observables beyond four dimensions is ambiguous. A jet algorithm typically clusters particles based on some angular distance between momenta. It is ambiguous whether - and how - the five- and six-dimensional components should be included in these definitions. Such choices can certainly have an impact on the contribution coming from the massification term $S_{b\overline{b}}\otimes\sigma$. However, the exact same ambiguity shows up in the massless cross section. There, the integrated subtraction terms for soft $b\overline{b}$-pairs likewise involve higher-dimensional momenta. Clearly, the physical cross section cannot depend on the choice of analytical continuation of four-dimensional observables, so the dependence on this choice must drop out in the combination of all terms.

There is a subtlety regarding the choice of analytical continuation of observables. It might appear as if one has complete freedom regarding how the fifth and sixth momentum components enter the observable definition, as long as the usual four-dimensional definition is recovered when the higher-dimensional components are set to zero. This is, however, not the case. Dimensional regularisation can be applied as long as the four-dimensional case is recovered in the limit $\varepsilon\to0$ up to higher-order terms in the $\varepsilon$-expansion. However, the angular integrations beyond four dimensions receive, in general, non-trivial contributions which do not vanish at $\varepsilon=0$. This can be seen in eq.~(88) of ref.~\cite{Czakon:2014oma}: while the integral over the fifth component is $\mathcal{O}(\varepsilon)$ for a trivial sixth component, the integral over the sixth component is $\mathcal{O}(\varepsilon^0)$. For physical cross sections, this is unproblematic, because there can only be a non-trivial dependence on the sixth component of the second unresolved momentum if the first has a non-zero fifth component. If this is the case, then the whole contribution is $\mathcal{O}(\varepsilon)$ on account of the non-zero fifth component of the first unresolved momentum. If the fifth component of the first unresolved momentum does vanish, then the integrand does not depend on the sixth component of the second unresolved momentum, leaving just the plus distribution from the phase space measure in eq.~(88) of ref.~\cite{Czakon:2014oma}, which integrates to zero. Because of this, it is sufficient to use four-dimensional momenta for the integration over the real emission phase space for contributions absent any $\varepsilon$-poles. However, if an explicit dependence on the $(d-4)$-dimensional components of momenta is introduced that involves more than just physical quantities, i.e.~scalar products of the $(d-4)$-dimensional parts of the momenta, then this can introduce an $\mathcal{O}(\varepsilon^0)$ difference w.r.t.~the four-dimensional case, yielding an incorrect result. Here, we will limit ourselves to the class of analytical continuations of four-dimensional observables which depend on higher-dimensional components of momenta only through scalar products of their $(d-4)$-dimensional parts, and in such a way that the four-dimensional definition is recovered when all these scalar products vanish. Such analytical continuations are guaranteed to only induce effects of $\mathcal{\varepsilon}$ beyond four dimensions. Within this class, one has complete freedom regarding the choice of analytical continuation and the dependence on this choice drops out in the final result, as discussed above.

Having dealt with the soft singularity, one important point remains: the massless contribution to $S_{b\overline{b}}$ also contains collinear singularities. When this contribution is omitted, these divergences are replaced with collinear singularities of the massive contribution in the UV. In practice, these are collinear singularities in the (integrated) subtraction term introduced to regulate the UV divergence. However, note that these collinear configurations cannot affect the flavour of jets. Indeed, the only singularities occur when the two $b$-quarks become collinear or when they additionally become collinear to one of the hard partons. Since the $b$-quarks must be collinear, they are necessarily clustered into the same jet and thus leave its flavour unchanged. This flavour-unaffected cross section also gets a contribution from $S_\emptyset$, which can be obtained by integrating the contribution of $S_{b\overline{b}}\otimes\sigma$ with the $b$-quarks removed with a minus sign. If we compute $S_\emptyset\otimes\sigma$ numerically in this way at the same time as $S_{b\overline{b}}\otimes\sigma$, it thus acts as a local subtraction term for the collinear singularities in $S_{b\overline{b}}\otimes\sigma$.

In practice, the implementation of the soft massification terms thus looks as follows at NNLO. A Born phase space configuration and the momenta $p_b$ and $p_{\overline{b}}$ are generated. The soft $b\overline{b}$-pair is inserted into the final state and the contributions from $S_{b\overline{b}}$ and its UV subtraction term are computed and added to the cross section. Then these same contributions are subtracted, after removing the soft quarks from the final state. The integrated subtraction term is treated analogously. This will simultaneously compute the contributions from $S_\emptyset$ and $S_{b\overline{b}}$, while regulating all singularities. Additionally, it is important that the computation of the massless cross section is kept fully differential in the momenta of the soft $b$-quark pair, including the five- and six-dimensional components where needed. The implementation of the sector-improved residue subtraction scheme \cite{Czakon:2010td,Czakon:2011ve,Czakon:2014oma,Czakon:2019tmo} in the C++ library \textsc{Stripper} has been extended to support our proposed solution to the jet flavour problem. \textsc{Stripper} has the advantage of having been designed with the aim of avoiding analytical integrations. In contrast to most implementations of subtraction schemes, the integrations over soft partons' angles are thus performed numerically, keeping the framework completely differential w.r.t.~such typically unobserved degrees of freedom, making the present extension and others like it essentially trivial.

While the focus here is on NNLO -- which is, after all, the only case of phenomenological relevance at this time -- it is apparent that our solution can easily be extended to higher orders. Since the IRC poles are theoretically cancelled locally (though the practical implementation used here is non-local), it is not affected by any of the problematic cases described in ref.~\cite{Caola:2023wpj}, as long as the higher-order extension is performed consistently. It should also be evident that the additional effort needed to implement our solution at a given order will be just a fraction of the effort needed to perform calculations at that order in the first place. The implementation at NNLO requires no analytical integration whatsoever, only requiring studying one of the simplest double-unresolved limits of amplitudes and some minor modifications to the subtraction scheme. At N$^3$LO, some one-loop integrations will need to be performed analytically, but the effort required to do so is minor compared to the effort needed to construct an N$^3$LO subtraction scheme and compute all its necessary ingredients. We therefore see no well-justified barrier to using this solution at higher orders.

\subsection{Logarithmic structure and perturbative convergence}
As with any factorisation, the factorisation of mass logs into the $S$ functions in principle introduces a new arbitrary renormalisation scale, we shall call it $\mu_S$, in analogy to $\mu_R$ for the strong coupling, $\mu_F$ for PDFs and $\mu_{Fr}$ for FFs. Our discussion so far has ignored this detail -- choosing effectively to set $\mu_S=\mu_R$ -- since the present work only considers fixed-order cross sections, where the dependence on this scale drops out exactly as long as the $\varepsilon$-poles cancel.\footnote{Which of course happens if the implementation is correct. Thus, generalising to $\mu_S\neq\mu_R$ and checking that the $\mu_S$-dependence drops out does not constitute an independent test of the correctness of the implementation.} That being said, if the soft mass logs are resummed, then the dependence on $\mu_S$ only cancels up to higher-order terms, as for all renormalisation scales.

Due to the more complicated nature of the $S$ functions, resummation of the soft mass logs cannot be done in a (semi-)analytical way as for couplings, FFs or PDFs. Realistically, a parton-shower-like approach will need to be used. Indeed, the flavour-changing soft logs are a type of non-global logarithm \cite{Dasgupta:2001sh,Banfi:2002hw}, which are notoriously tricky to resum and were first resummed in ref.~\cite{Dasgupta:2001sh} using a numerical Monte Carlo approach. We will not spend a lot of time musing about how the resummation for the present case might be done in this work. Instead, we will argue that resummation is not necessary, at least for the foreseeable future. To this end, we consider the structure of the soft mass logs through the first few orders in perturbation theory. Obviously, the leading-order and NLO contributions are trivial. The first non-trivial order is NNLO, where a single log appears, corresponding to the simple $\varepsilon$-pole caused by the double-soft limit of a massless quark-anti-quark pair. The scaling through NNLO is thus $\alpha_s^2\ln(m_b^2/\mu_S^2)$, where $\mu_S$ should be set to the hard scale of the process. Due to the extra suppression by one power of $\alpha_s$ compared to e.g.~collinear mass logs, we expect that the soft mass logs at NNLO will be quite small. This will indeed be confirmed numerically in the next section.

\begin{figure}[t]
	\centering
	\includegraphics[width=0.49\textwidth]{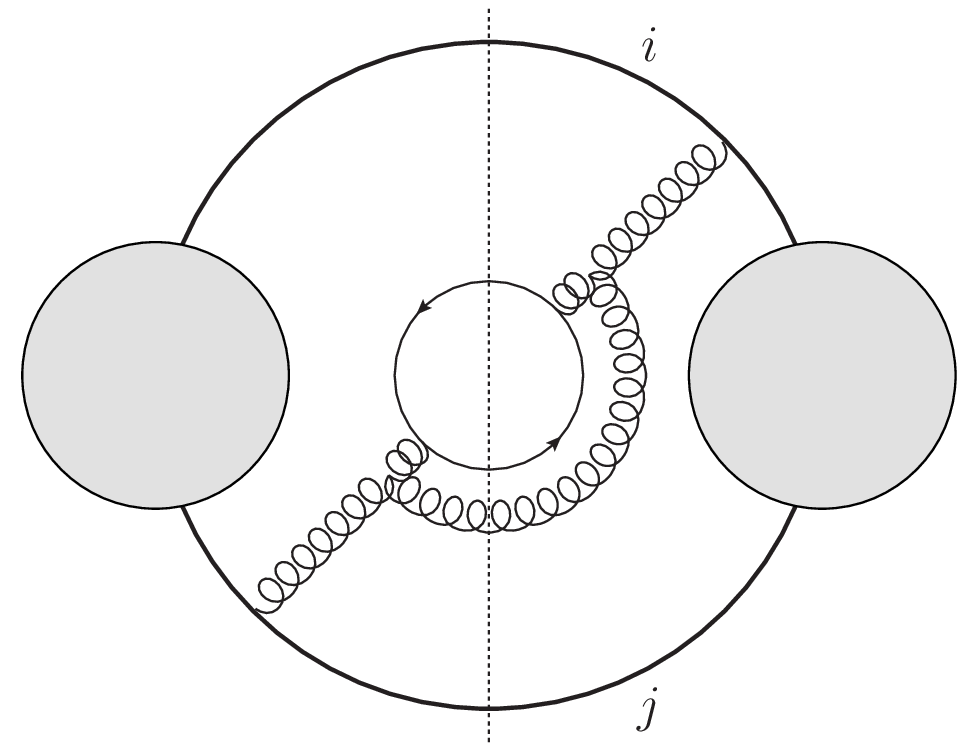}
	\caption{An example of an N$^3$LO contribution to $S_{b\overline{b}}$ which generates a non-cancelling double log.}
	\label{fig:N3LOContribution}
\end{figure}

In general, the behaviour at N$^n$LO is $\alpha_s^n\ln^{n-1}(m_b^2/\mu_S^2)$.\footnote{This behaviour is derived here under the assumption of QCD colour coherence. It is known \cite{Catani:2011st,Forshaw:2012bi,Schwartz:2017nmr,Henn:2024qjq,Guan:2024hlf} that QCD colour coherence is violated at sufficiently high orders for processes with hadronic initial states. This can lead to double logarithmic effects \cite{Forshaw:2008cq,Becher:2021zkk,Becher:2023mtx,Dasgupta:2025cgl}. We leave an analysis of the impact of colour coherence violating effects on flavoured jet production for future work, but note that even in the complete absence of colour coherence, the behaviour would at most be $\alpha_s^n\ln^{2n-3}(m_b^2/\mu_S^2)$, significantly suppressed relative to the naive expectation.} Assuming $\alpha_s\ln(m_b^2/\mu_S^2)$ to be $\mathcal{O}(1)$, the soft mass logs at each consecutive order will give a contribution similar in size to the one at NNLO. While in theory this is the classic situation requiring resummation, in practice the resummation may not be relevant until a sufficiently high order is reached. This can be seen by considering the leading logarithmic contribution at a given order. An example diagram at N$^3$LO yielding a (non-cancelling) double logarithm is shown in fig.~\ref{fig:N3LOContribution}. The double logarithm originates from iterating soft limits: first, the $b\overline{b}$-pair is taken to be soft, and then the gluon can exhibit a soft singularity as well. In general, the leading logarithmic contribution at any order receives contributions from such iterated limits. At N$^n$LO, one can obtain a leading-log contribution by first emitting $n-1$ soft gluons, each emitted by the previous one and each parametrically softer than the previous, and then splitting the final gluon into a wide-angle $b\overline{b}$-pair. Each consecutive gluon introduces a factor $\alpha_s\ln(m_b^2/\mu_S^2)$, except for the last gluon splitting into a $b\overline{b}$-pair, which only contributes $\alpha_s^2\ln(m_b^2/\mu_S^2)$. This gives the $\alpha_s^n\ln^{n-1}(m_b^2/\mu_S^2)$ scaling mentioned before. This also highlights how these logs might be resummed. One defines generalisations of the $S$ functions to remain differential in all soft partons (i.e.~introduces $S_g$ at NLO, etc.) and then iteratively adds soft radiation according to these $S$ functions in the style of a parton shower. In practice, this (leading-logarithmic) resummation will be significantly simpler than standard parton showers, since we are only interested in iterating strict soft limits, without momentum recoil, etc. Again, we do not wish to go into further detail about this here.

The point is that the leading log at N$^n$LO will thus, schematically, always look like $S_{b\overline{b}}\times S_g^{(n-2)}$, which is resummed into $S_{b\overline{b}}\times\mathcal{O}(1)$. Thus if the contribution from the NNLO soft mass logs is small, then the entire leading-log contribution will be of a similar magnitude. Thus, if we find that the NNLO soft mass logs are significantly smaller than the other theory uncertainties, then whether or not the logs are resummed is inconsequential. The resummation will only need to be performed once the total contribution from soft mass logs up to a given order starts reaching a magnitude similar to the scale uncertainties (or other uncertainties) at that order. Considering how small the contribution is at NNLO (as we will see in section \ref{sec:Pheno}), resummation is definitely not necessary at that order. Additionally, once such a precision has been reached that one is sensitive to these effects, power corrections in the mass will be relevant as well, in which case the massification discussed here is irrelevant, since one has to work with massive quarks anyway. We will show this explicitly in section \ref{sec:Pheno}.

\subsection{Renormalisation and non-perturbative flavour}
As discussed before, the $\varepsilon$-poles produced by the $S$ functions are analogous to the $\varepsilon$-poles present in perturbative PDFs and FFs. In those cases, they are typically removed by collinearly renormalising those objects. Then, collinear renormalisation counterterms are introduced in massless cross section computations to cancel the homologous divergences seen there. In effect, one is not truly renormalising in the traditional sense, but merely shifting terms around to make two separately divergent quantities separately finite. One could similarly perform a `soft renormalisation' of the $S$ functions and introduce by hand `soft renormalisation counterterms' to the massless cross section. We spare ourselves this exercise here, since it is pointless for purely perturbative fixed-order quantities.

That being said, it is perfectly valid to define non-perturbative versions of the $S$ functions, just like one defines (non-perturbative) fragmentation functions for hadronisation. This would introduce a set of non-perturbative $S$ functions, which can be softly-renormalised to render e.g.~strange-jet cross sections finite. They would be subject to renormalisation group equations, which however would have to be solved using parton-shower techniques, as mentioned above. As always, they must be fitted to data, but are universal, i.e.~process-independent. Treating any quark as `flavoured', one can even use such functions to define IRC-safe quark and gluon flavour modulo-2 jets. While an interesting concept, we refrain from going into more detail here, as a proper exploration of this topic would be beyond the scope of this introductory work.

We will now present some tests of the implementation of our scheme in \textsc{Stripper} and consider its phenomenology.

\section{Numerical checks}\label{sec:Checks}

To test our idea and its implementation in \textsc{Stripper}, we first consider the single-inclusive $b$-jet cross section for the 13 TeV LHC, where partons are clustered using the anti-$k_T$ algorithm with $R=0.4$ and $b$-jets are defined using the flavour modulo-2 scheme. We will consider the $p_T$ spectrum from $25$ GeV to $1600$ GeV and require $\lvert y_{b\textrm{-jet}}\rvert < 2$. Our default choice for the parameters of the UV subtraction term, cf.~eq.~\eqref{eq:UVsubtraction}, will be $A=3$, $B=2$ and $R=m_b$.\footnote{The justification for this choice is as follows. Choosing $B=2$ seems the most natural, while choosing $A=3$ matches the behaviour of the $S$ function for small three-momenta. Natural choices of $R$ should be proportional to $m_b$. Which one of $R=m_b$ and $R=2m_b$ is more natural is less obvious, but in the end these choices only affect numerical convergence and even there their impact should not be huge, provided they are chosen within reason.} For the analytic continuation of four-dimensional observables to five or six dimensions, our default choice will be to define azimuthal angles in terms of $p_1$ and $p_2$ and rapidities in terms of $p_0$ and $p_3$ using the usual definitions. Explicitly:
\begin{equation}
	y(p) = \frac{1}{2}\ln\bigg(\frac{p_0+p_3}{p_0-p_3}\bigg)\;,\;\;\;\;	\Delta\phi(p_a,p_b) = \acos\Bigg(\frac{p_{a,1}p_{b,1}+p_{a,2}p_{b,2}}{\sqrt{\big(p_{a,1}^2+p_{a,2}^2\big)\big(p_{b,1}^2+p_{b,2}^2\big)}}\Bigg)\;.
\end{equation}
In practice, these are the only observables of five- and six-dimensional momenta that enter the calculation.\footnote{Note that the choice of analytic continuation of (infinitesimal) $p_T$ has no impact on anti-$k_T$ jets.}

We choose the central scale $\mu_R=\mu_F=p_T$, where $p_T$ is the transverse momentum of the jet being binned. We use the NNLO NNPDF3.1 PDF+$\alpha_s$ set \cite{NNPDF:2017mvq}, taken from the LHAPDF interface \cite{Buckley:2014ana}, fixing $m_b=4.92$ GeV to be consistent with that set.

\begin{figure}[t]
	\centering
	\includegraphics[width=0.49\textwidth]{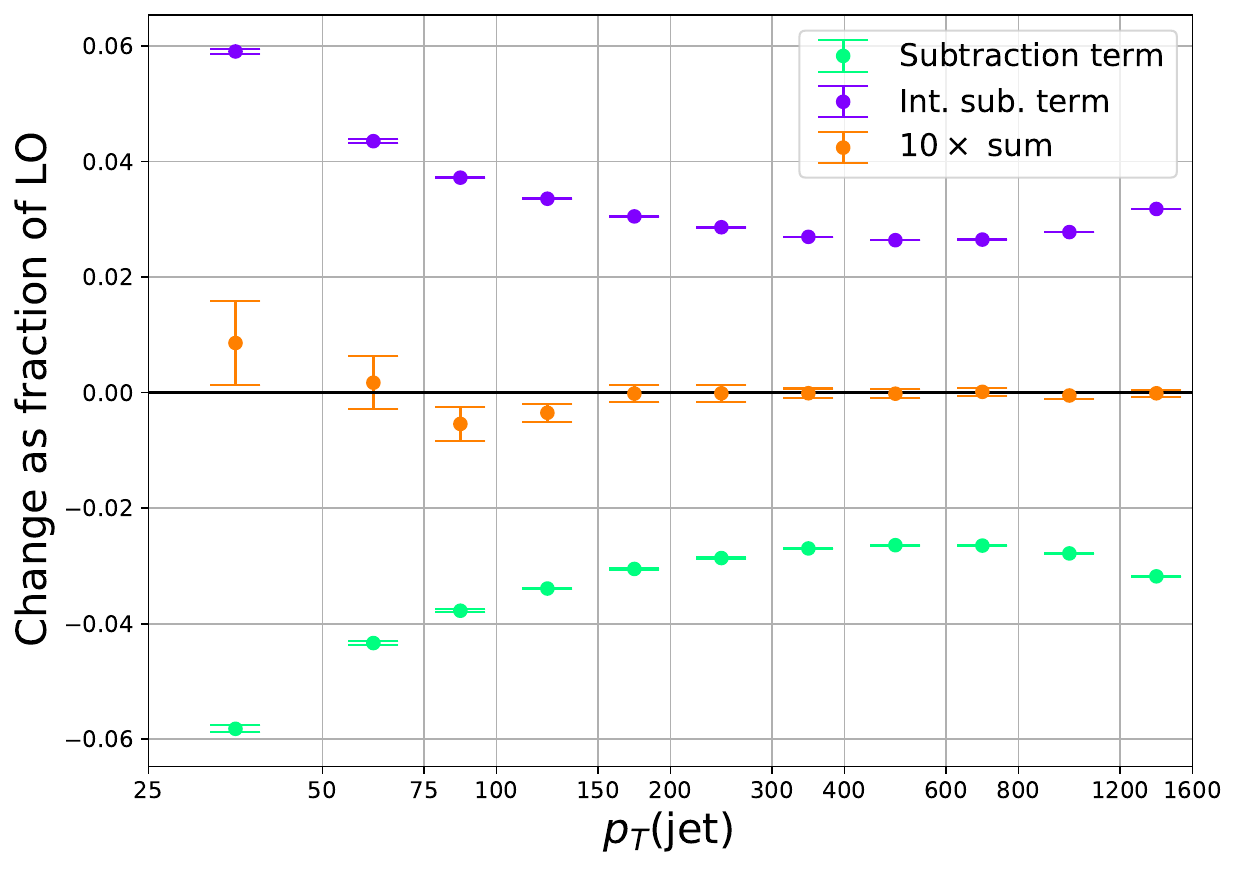}
	\includegraphics[width=0.49\textwidth]{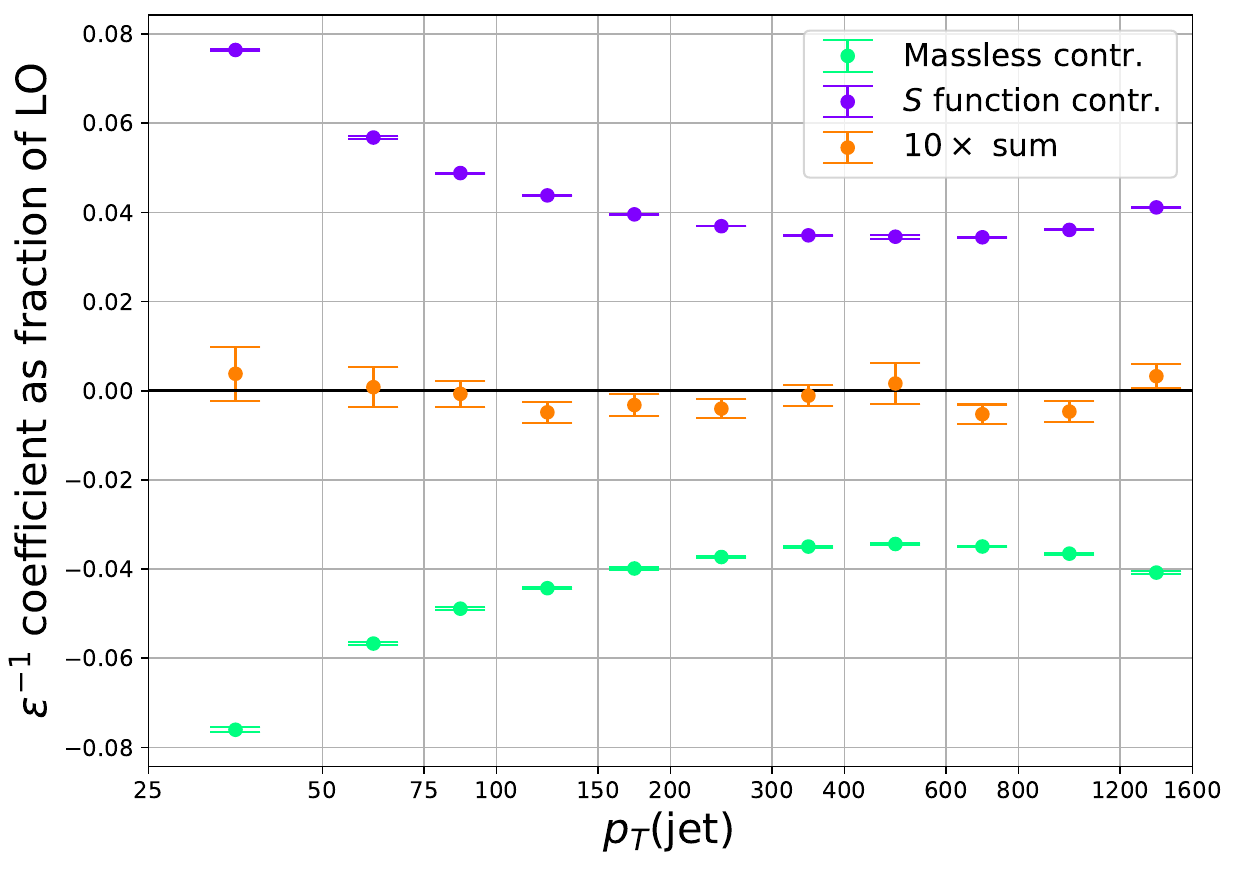}
	\includegraphics[width=0.49\textwidth]{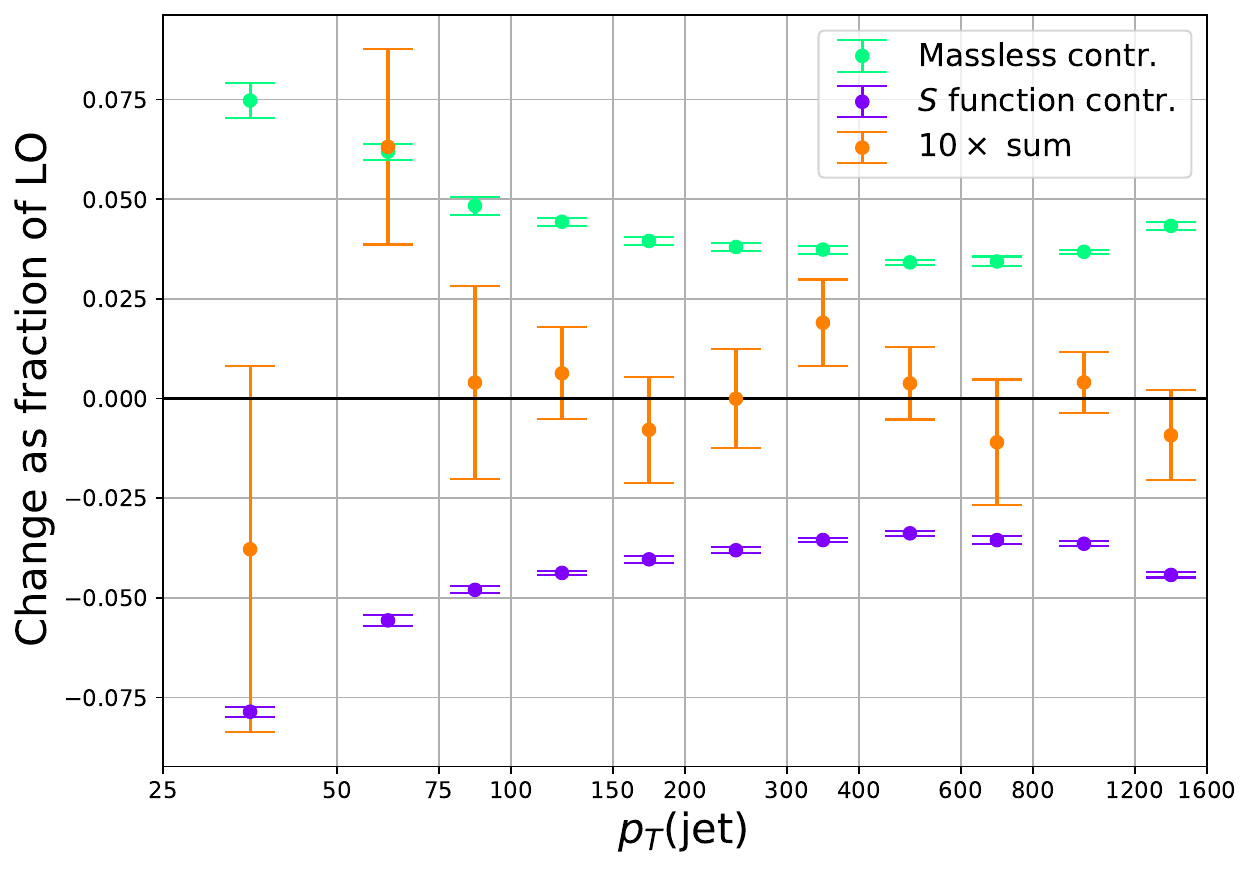}
	\caption{Top left: the effect of changing the parameters $A$, $B$ and $R$ as described in the text on the subtraction term contribution (green), the integrated subtraction term contribution (purple) and their sum (orange, magnified 10$\times$). Shown are the effects at NNLO as fractions of the LO cross section. Top right: the $\varepsilon^{-1}$ contribution of the massless cross section (green), the $S$ function contribution (purple) and their sum (orange, magnified 10$\times$) as fractions of the LO cross section. Bottom: the effect of changing the $d$-dimensional definitions of $y$ and $\Delta\phi$ as described in the text on the massless cross section (green), the $S$ function contribution (purple) and their sum (orange, magnified 10$\times$) as fractions of the LO cross section.}
	\label{fig:ABR4vD}
\end{figure}

We present four tests of our implementation. The first test is to make sure the $S$ function contribution is independent of the choice of the parameters $A$, $B$ and $R$ in eq.~\eqref{eq:UVsubtraction}. We can compute the four-dimensional finite part and the $d$-dimensional integrated subtraction term separately for two different choices of the parameters, verify that the individual contributions change and then check that their sum does not change. To perform this test, we compute the difference between our default choice and the choice $A=0$, $B=1$ and $R=2m_b$. We explicitly present the result for the sum over all partonic configurations in the left panel of fig.~\ref{fig:ABR4vD}. Clearly, the difference between the two parameter choices is non-zero for both the four-dimensional contribution and the integrated subtraction term, while this dependence cancels in their sum. It was checked that this also holds for each $2\to2$ partonic Born configuration separately. This test has thus been passed.

The second test is to make sure that the poles in $\varepsilon$ cancel between the massless jet cross section and the contribution from the $S$ functions. The central panel of fig.~\ref{fig:ABR4vD} shows this for the full NNLO result. We can clearly see that the massless cross section contains a leftover simple pole in $\varepsilon$ - this is precisely the result of the non-IRC safety of the standard definition of jet flavour.\footnote{Note that while in previous versions of \textsc{Stripper}, soft wide-angle flavour-changing emissions were not subtracted in a way that preserves all information on the soft partons, leading to non-integrable integrands, this updated version of \textsc{Stripper} fixes this limitation, fully locally subtracting the singularity and correctly adding the corresponding $\varepsilon$-pole fully differentially. As a result, this new version can be used to numerically test the safety of jet flavour definitions through NNLO by confirming the cancellation of $\varepsilon$-poles. Alternatively - and more efficiently - it can be checked that the $S$ function contribution vanishes identically, i.e.~does not contribute for any phase space point.} Adding the $S$ function contribution cancels this leftover pole, rendering the whole calculation finite. Again, while only the results for the sum over all channels are shown here, it was checked that the poles also cancel for each combination of initial-state partons separately. Thus, this test has also been passed. Pole cancellation was also verified for the other processes discussed in the remainder of this work.

The third test is to verify that the dependence on how observables are defined beyond four dimensions drops out in the sum over contributions. In this case, we calculate the difference between results obtained using our default choice and those obtained using the following definitions of rapidity and $\Delta\phi$:
\begin{equation}
y(p) = \frac{1}{2}\ln\bigg(\frac{\sqrt{p^2+p_1^2+p_2^2+p_3^2}+p_3}{\sqrt{p^2+p_1^2+p_2^2+p_3^2}-p_3}\bigg) = \frac{1}{2}\ln\bigg(\frac{\sqrt{p_0^2-p_4^2-p_5^2}+p_3}{\sqrt{p_0^2-p_4^2-p_5^2}-p_3}\bigg)\;,
\end{equation}
\begin{equation}
\Delta\phi(p_a,p_b) = \acos\Bigg(\frac{\sum_{i\in\{1,2,4,5\}}p_{a,i}p_{b,i}}{\sqrt{\big(\sum_{i\in\{1,2,4,5\}}p_{a,i}^2\big)\big(\sum_{i\in\{1,2,4,5\}}p_{b,i}^2\big)}}\Bigg)\;.
\end{equation}
The result of this test is shown in the right panel of fig.~\ref{fig:ABR4vD}. Again, the massless cross section clearly depends on the choice of analytic continuation of observables. The dependence of the $S$ function contribution is equal and opposite, so the full cross section is independent of these choices. As before, only the results for the sum over all $2\to2$ partonic Born configurations are shown here, but it was checked that the dependence drops out also for each Born configuration separately. Therefore, this third test has also been passed.

Note that these first three tests are mathematically equivalent, since the terms being added to yield zero are always proportional to the non-cancelling $\varepsilon$-pole of the massless cross section. However, they each test completely independent parts of the practical implementation. In fact, combined with the final test, they cover all aspects of the implementation.

The final test is obvious: the massified cross section should match the massive one up to power corrections. In order to eliminate the effect of power corrections and maximise the size of the quasi-soft logarithms, this test should be performed for a very small value of the massive quark mass. However, this will lead to very large cancellations between different contributions to the massive calculation: the collinear (and part of the soft) logarithms cancel in the final result, but are present in the individual contributions to the cross section. This means that the individual cross section contributions need to be determined with unusually high numerical precision, i.e.~a small Monte Carlo uncertainty. While not impossible, it would be unnecessarily computationally expensive to perform this test for an LHC cross section. For the sake of the environment, we will therefore instead consider the cross section for producing top-jets at a hypothetical 1 TeV electron-positron collider, setting $m_t = 1$ GeV and including only the QED contribution for simplicity. This process actually constitutes a very rare exception which cannot be handled by the original formulation of the sector-improved residue subtraction scheme. To solve this, we have extended the scheme to cover such edge cases, as described in appendix \ref{sec:appendix}. The small mass also makes the cross section hard to integrate using Monte Carlo techniques due to the presence of quasi-singularities. This is also solved by the strategy described in appendix \ref{sec:appendix}.

\begin{figure}[t]
	\centering
	\includegraphics[width=0.49\textwidth]{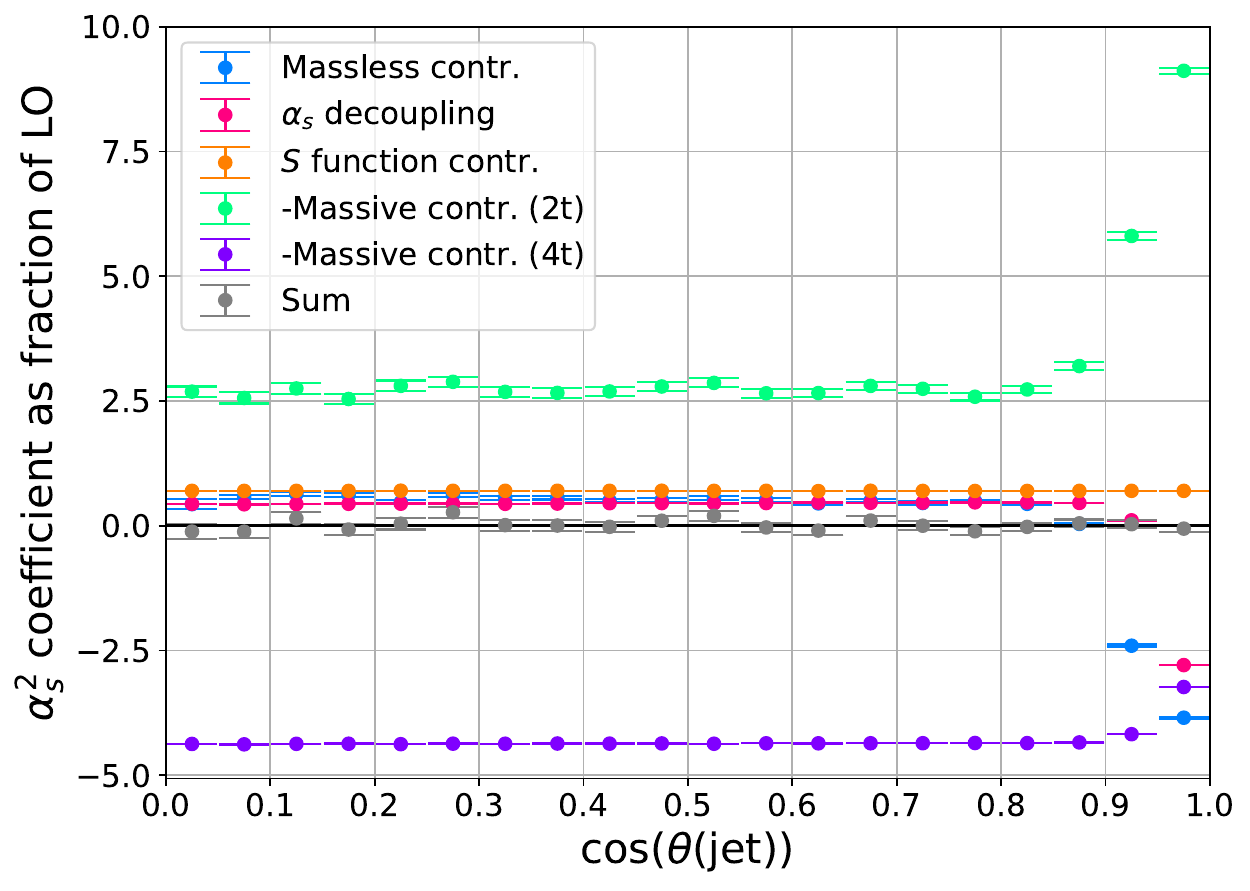}
	\includegraphics[width=0.49\textwidth]{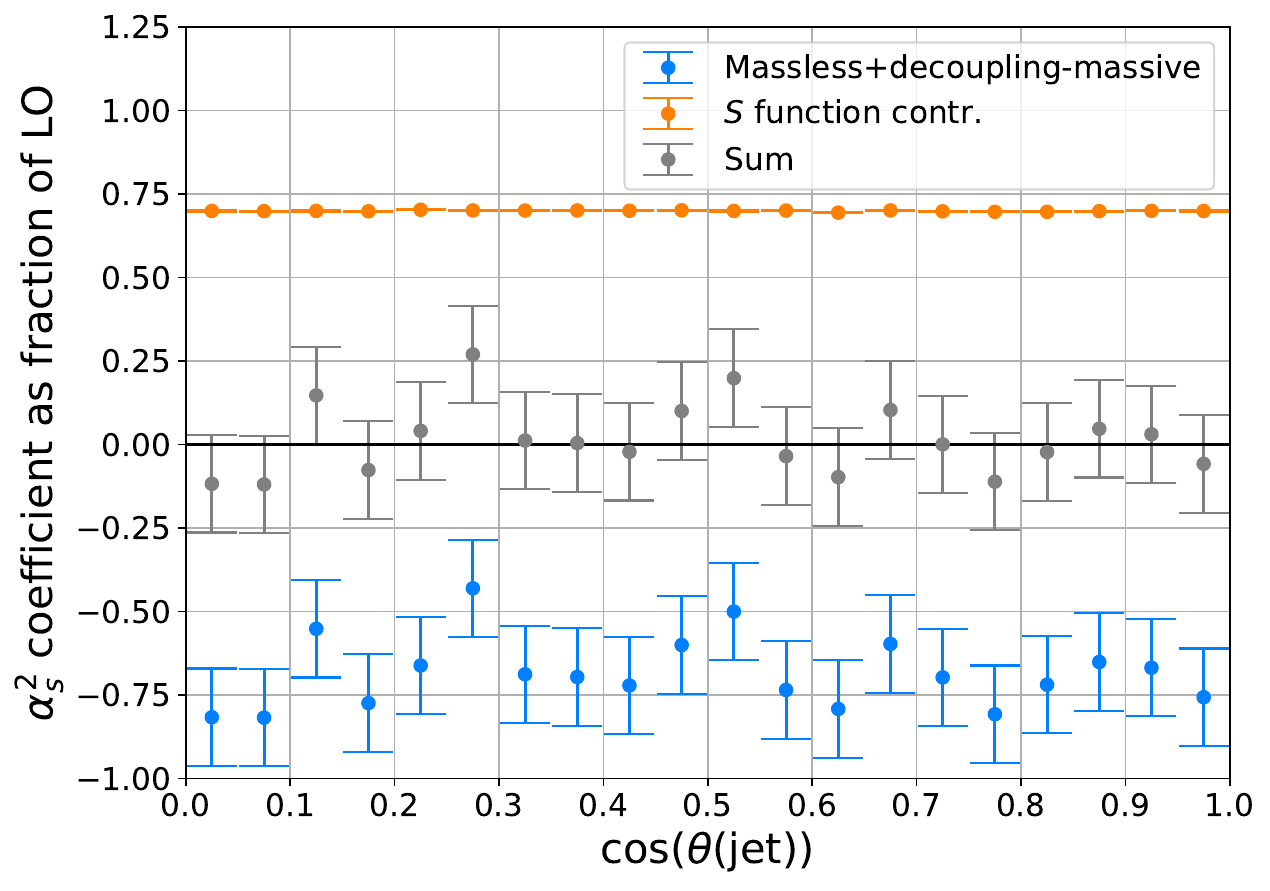}
	\caption{Left: shown are the different NNLO contributions to the LP hypothetical $e^+e^-$ to top-jet cross section described in the text, namely the purely massless contribution (blue), the $\alpha_s$ decoupling contribution (pink) and the $S$ function contribution (orange), as well as the NNLO contributions to the fully massive cross section, distinguishing between final states with two (green) and four (purple) top quarks and plotted with an additional minus sign. The sum of all of these contributions is shown in grey. Right: the $S$ function contribution to the LP cross section (orange), the sum over all contributions shown in the left panel except for the $S$ function contribution (blue) and the sum of both of these (grey).}
	\label{fig:EETT}
\end{figure}

We cluster jets using the Durham algorithm with $y=10^{-2}$ and study the distribution of the zenith angle of the highest-energy top-jet. The left panel of fig.~\ref{fig:EETT} shows the different NNLO contributions to the leading power cross section: the purely massless contribution, the $\alpha_s$ decoupling contribution and the $S$ function contribution. Also shown are the NNLO contributions to the fully massive cross section, distinguishing between final states with two and four top quarks and plotted with an additional minus sign. If the LP contributions have all been computed correctly, then they should - up to negligible power corrections - add up to the fully massive cross section. Thus, adding all of the plotted contributions listed above should yield zero. Note that all plotted contributions have a different shape, requiring a non-trivial interplay in order for them to sum up to zero. Their sum is shown in grey, and is indeed consistent with zero.

This can be seen more clearly in the right panel of fig.~\ref{fig:EETT}. Shown there are the $S$ function contribution in orange and the sum of all other contributions in blue. Clearly, both are constant w.r.t.~the zenith angle in units of the LO cross section and their sum is zero. Among those presented here, this is the most powerful test of the implementation of our idea - and indeed of the concept itself. And it, too, has been passed.

Having confirmed that everything is working as expected, we now turn to a pair of phenomenological applications. 

\section{Phenomenology}\label{sec:Pheno}

\subsection{$b$-jet production in association with a $Z$-boson at the LHC}

We will first consider the transverse momentum spectrum of the leading $b$-jet produced in association with a $Z$-boson (decaying to a pair of muons) at the 13 TeV LHC. This process was first considered through NNLO in ref.~\cite{Gauld:2020deh} and was studied extensively in ref.~\cite{Behring:2025ilo} using several different IRC-safe definitions of jet flavour. To facilitate a direct comparison with the results of that study, we will use an identical setup for our calculation. This means we will work with $n_f=5$ and apply the following cuts:\footnote{Note that the lower invariant mass cut is erroneously listed as $77$ GeV in ref.~\cite{Behring:2025ilo} instead of $71$ GeV.}
\begin{itemize}
	\item Both muons must satisfy $p_T(\mu) > 20$ GeV and $|y(\mu)| < 2.4$
	\item $p_T(\mu\mu) > 20$ GeV and $71$ GeV $< m(\mu\mu) < 111$ GeV
	\item At least one flavour-modulo-2 $b$-tagged $R=0.5$ anti-$k_T$ jet with $p_T(b$-jet$) > 30$ GeV,  $|y(b$-jet$)| < 2.4$ and $\Delta R(\mu,b$-jet$) > 0.5$ w.r.t.~both muons.
\end{itemize}
Furthermore, we will use the PDF4LHC21 PDF and $\alpha_s$ set \cite{PDF4LHCWorkingGroup:2022cjn} and use the $G_F$ scheme for the electroweak coupling, with $m_W=80.352$ GeV, $m_Z=91.1535$ GeV, $\Gamma_W=2.084$ GeV, $\Gamma_Z=2.4943$ GeV and $G_F=1.16638\cdot10^{-5}$ GeV$^{-2}$. As before, we set $m_b=4.92$ GeV. We will perform a 7-point scale variation around $\mu_R=\mu_F=\sqrt{m^2(\mu\mu)+p_T^2(\mu\mu)}$. The only difference between our calculation and the calculations of ref.~\cite{Behring:2025ilo} is the approach to flavour tagging.

We will particularly focus on comparing our results obtained using anti-$k_T$ to results obtained using the CMP jet algorithm \cite{Czakon:2022wam}.\footnote{The original formulation of the CMP algorithm had some flaws, which were pointed out in ref.~\cite{Caola:2023wpj}. In that work, a modification of the algorithm was suggested to resolve these issues. Later, the algorithm was further modified in ref.~\cite{Behring:2025ilo}. We will employ this latest version of the CMP algorithm throughout this work.} This algorithm is identical to anti-$k_T$, except it modifies the distance measure for pairs of pseudojets with non-zero net flavour numbers that are equal in magnitude but opposite in sign as follows:
\begin{equation}
	\begin{gathered}
		d_{ij}^\text{flavoured} = d_{ij}^\text{anti-$k_T$}\times\bigg[1-\theta(1-\kappa_{ij})\cos(\frac{\pi}{2}\kappa_{ij})\bigg]\;,\\
		\kappa_{ij} = \frac{1}{a}\frac{p_{T,i}^2+p_{T,j}^2}{p_{T,\text{max}}^2}\sqrt{\frac{\Omega_{ij}^2}{\Delta R_{ij}^2}}\;,\;\;\;\;\;
		\Omega_{ij}^2 = 2\bigg[\frac{1}{\omega^2}\big(\cosh(\omega\Delta y_{ij})-1\big)+\big(\cos(\Delta \phi_{ij})-1\big)\bigg]\;,
	\end{gathered}
\end{equation}
where $a$ is a new free parameter, $p_{T,\text{max}}$ is the highest transverse momentum among the pseudojets at a given point in the clustering procedure and $\omega>1$ is another free parameter. We will choose $\omega=2$ throughout this work and fix $a=0.1$ in this subsection.

\begin{figure}[t]
	\centering
	\includegraphics[width=0.49\textwidth]{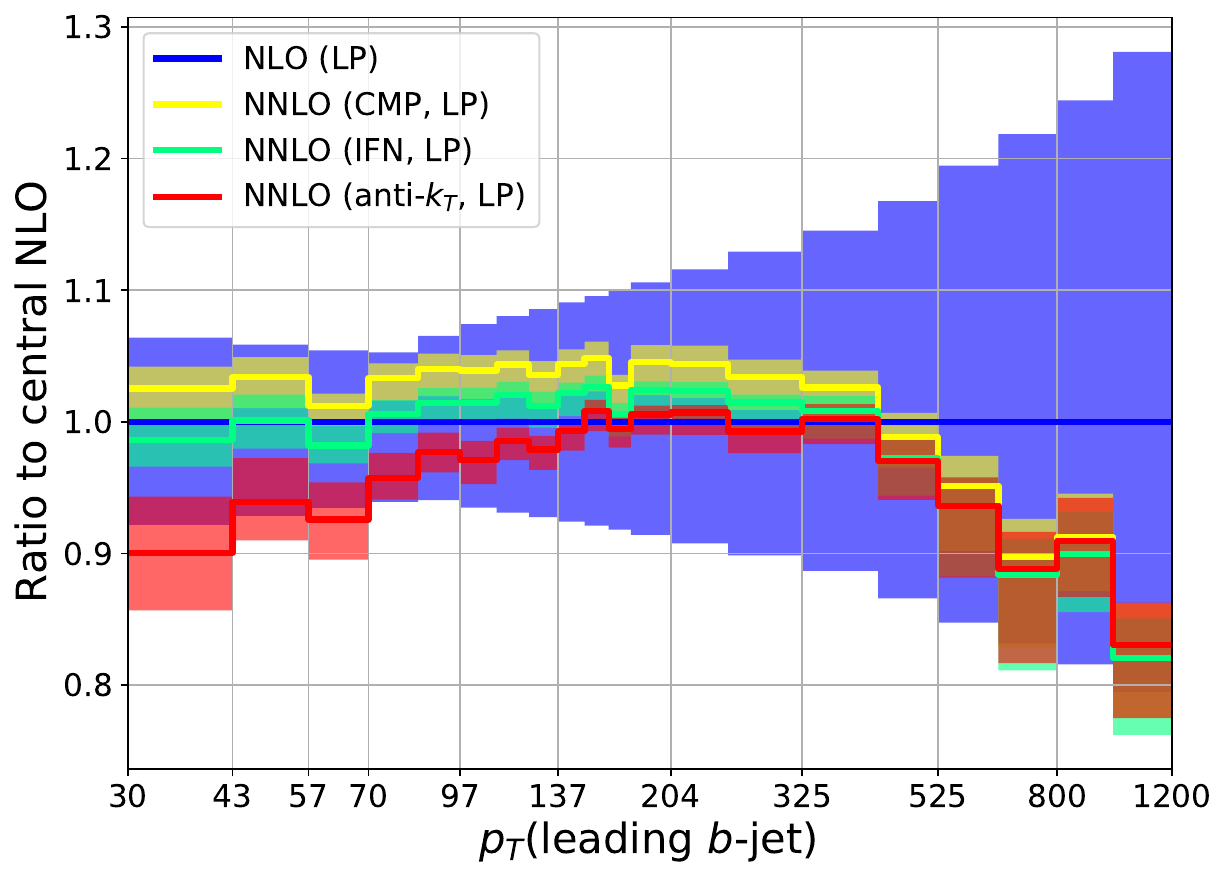}
	\includegraphics[width=0.49\textwidth]{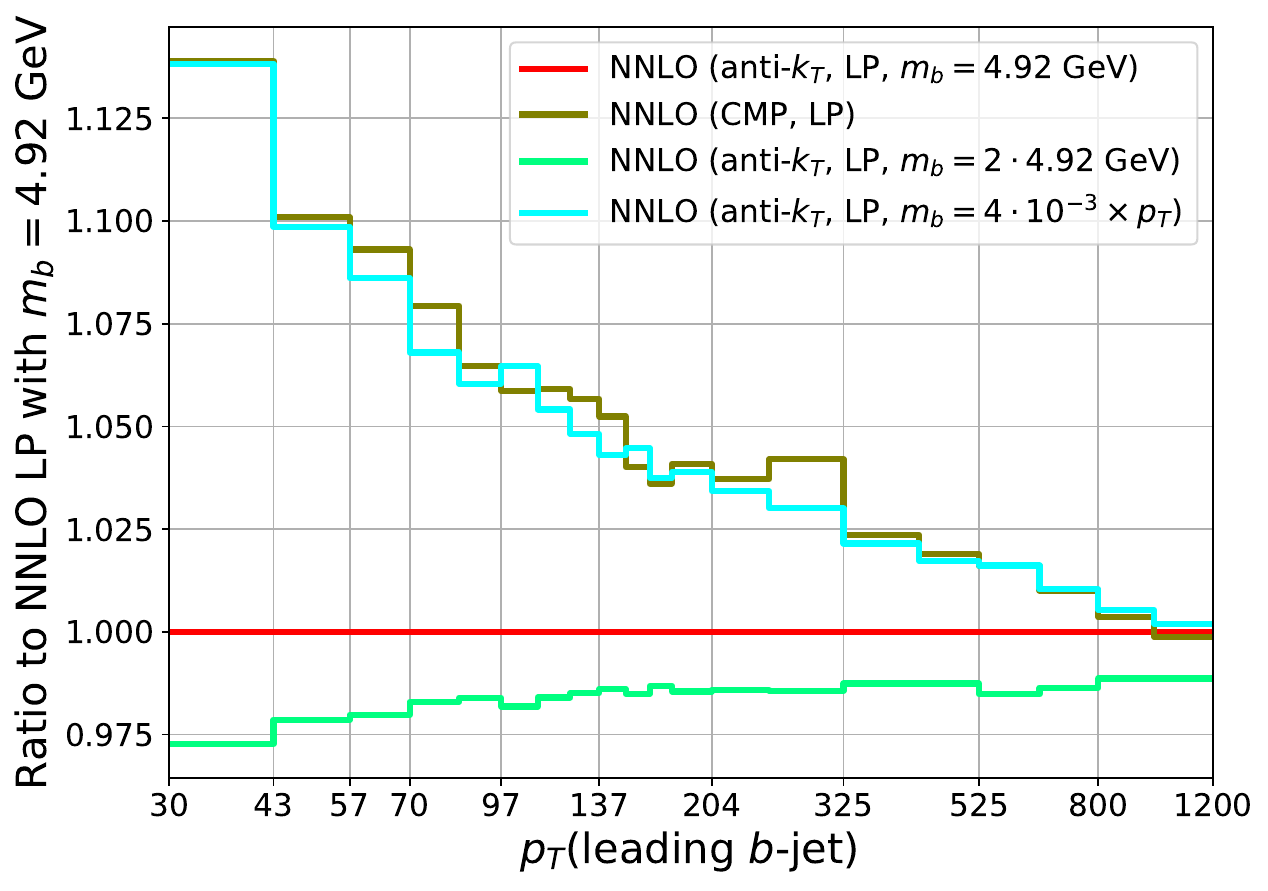}
	\caption{Left: the leading-$b$-jet transverse momentum spectrum for $Z+b$-jet production at NLO (blue) and NNLO, using the anti-$k_T$ (red), CMP (yellow) and IFN (lime) jet algorithms. The results obtained using CMP and IFN were taken from ref.~\cite{Behring:2025ilo}. Right: a comparison of the NNLO predictions obtained using anti-$k_T$ (red) and CMP (brown) at LP, and the impact of doubling the $b$-quark mass (green). Also shown is a result using anti-$k_T$ with a small, dynamical mass (cyan).}
	\label{fig:JFTppZbRene}
\end{figure}

The results are shown in the left panel of fig.~\ref{fig:JFTppZbRene}. We explicitly compare results obtained using anti-$k_T$ within our approach to results obtained using the CMP and IFN \cite{Caola:2023wpj} algorithms, which together give a good impression of the general spread of results produced by the different algorithms tested in ref.~\cite{Behring:2025ilo}. At high transverse momentum, our approach yields results very similar to those of the other two algorithms. For lower transverse momenta, however, our approach yields a cross section that is about 10\% lower than the other algorithms. This is noteworthy for two reasons. First of all, all the algorithms tested in ref.~\cite{Behring:2025ilo} yielded results that lie within just a few percent of each other at low $p_T$. Second of all, one would naively expect flavoured algorithms to yield smaller cross sections than anti-$k_T$, since the use of flavoured algorithms reduces the amount of flavoured jets and thus the amount of events that pass the phase space cuts. Here, for all but the highest transverse momenta, the opposite is observed.

One might suspect that these unexpected properties are the result of the `large' logarithm $\ln p_T^2/m_b^2$ introduced by the $S$ function contribution. However, the logarithm is not to blame, as can be concluded from the right panel of fig.~\ref{fig:JFTppZbRene}, which directly compares the NNLO results obtained using CMP, using our approach and the result obtained using our approach when increasing the $b$-quark mass by a factor of two.\footnote{We only change the $b$-quark mass in the $S$ function contribution, since this is the only source of non-resummed (and therefore potentially problematic) logarithms of the mass.} First of all, the effect is very small - at most only about 2\%. Since the relevant `hard' scale in the first $p_T$ bin (where the largest deviation relative to the other flavour definitions is observed) is $R\cdot p_T \sim 15$ GeV, the `large' logarithm would be eliminated if $m_b\sim15$ GeV. Since a factor of two already yields a mass reasonably close to that, it seems unlikely that the `large' logarithm could give a contribution beyond just a few percent. More importantly, increasing the $b$-quark mass - and thus reducing the `large' logarithm - actually decreases the cross section, further increasing the gap between our approach and those tested in ref.~\cite{Behring:2025ilo}.

This now suggests that one might be able to reproduce the results of the other methods by using a smaller $b$-quark mass, increasing the `large' logarithm. Most of the flavoured jet algorithms in the literature effectively introduce a cutoff energy, below which clustering two flavoured jets is favoured. E.g.~in the case of CMP, clustering two flavoured jets is increasingly favoured for transverse momenta below roughly $\sqrt{a}\cdot p_{T,\text{max}}$. This puts an effective cutoff on the energy of soft flavoured quark pairs, regulating the singularity and introducing the logarithm $\ln a$. If one now imagines taking $a$ to be incredibly small, then the cross section is obviously sensitive to unphysical regions of the phase space, where the flavoured quarks are allowed to be softer than they could physically be, due to their finite mass. In the physical cross section, which our approach reproduces up to power corrections, the finite $b$-quark mass sets the effective energy cutoff. Any flavoured jet algorithm should thus ensure that the effective energy cutoff it implements is never below this physical effective energy cutoff. This is especially important at low $p_T$, where probing unphysically small scales can easily lead to inaccurate cross sections, as has been known for e.g.~inclusive heavy quark production for a long time \cite{Cacciari:1998it}.

The cyan curve in fig.~\ref{fig:JFTppZbRene} shows the cross section obtained when using anti-$k_T$ with a dynamical $b$-quark mass of $4\cdot10^{-3}$ times the $p_T$ of the $b$-jet. For this dynamical choice of the mass, it would appear that our approach can reproduce the CMP results, suggesting this is the `effective' $b$-quark mass implemented by that algorithm in this case for $a=0.1$. This would certainly qualify as being too low for transverse momenta below 1 TeV. However, it is worth remembering that all of these results - including the literature results we are comparing to - are only accurate up to power corrections in the $b$-quark mass. It is, in principle, possible that different approaches lead to wildly different power corrections, which could also explain the observed differences.

To test this, we can compare massless results to massive ones and study the dependence of the transverse momentum spectrum on the $b$-quark mass. To this end we will consider a slightly simplified setup to aid with numerical convergence. This time, the $Z$-boson will be on-shell and we will apply the following cuts:
\begin{itemize}
	\item $p_T(Z) > 20$ GeV and $|y(Z)| < 2.4$
	\item At least one $b$-tagged jet with $p_T(b$-jet$) > 30$ GeV and $|y(b$-jet$)| < 2.4$.
\end{itemize}
We will use the $n_f=4$ variant of PDF4LHC21 \cite{PDF4LHCWorkingGroup:2022cjn}. For the massless results, we will include - in addition to the $S$ function - the PDF and $\alpha_s$ \cite{Wetzel:1981qg,Bernreuther:1981sg,Bernreuther:1983zp,Bernreuther:1983ds,Larin:1994va} threshold matching conditions through fixed-order NNLO, as previously implemented in \textsc{Stripper} in ref.~\cite{Czakon:2024tjr}. This ensures that our massless results are LP accurate. Note that, since we are working with $n_f=4$, the leading contribution to the cross section is one order higher than before: whereas before, the Born process corresponded to the final state $Z+b/\overline{b}$, this is no longer allowed. The leading contribution thus corresponds to the final state $Z+b+\overline{b}$. We will thus only need to consider NLO massive cross sections. We will label the LP cross sections in this case as `NLO', since it corresponds to the LP approximation of the NLO massive cross section, but its singularity structure is that of an NNLO cross section, with collinear singularities already present at `LO'. Crucially, this also means that the problematic double-soft $b$-quark-pair singularity is present at `NLO'.

\begin{figure}[t]
	\centering
	\includegraphics[width=0.49\textwidth]{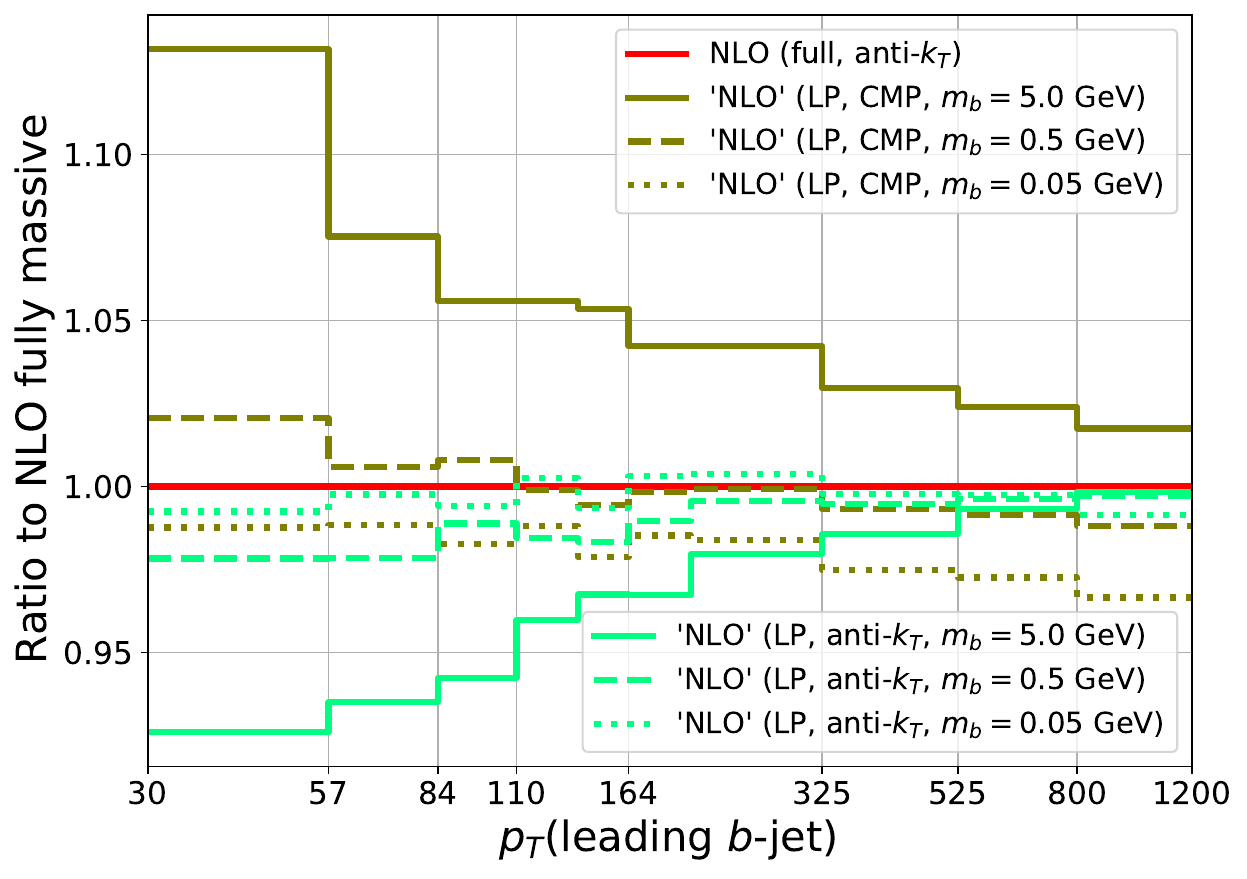}
	\includegraphics[width=0.49\textwidth]{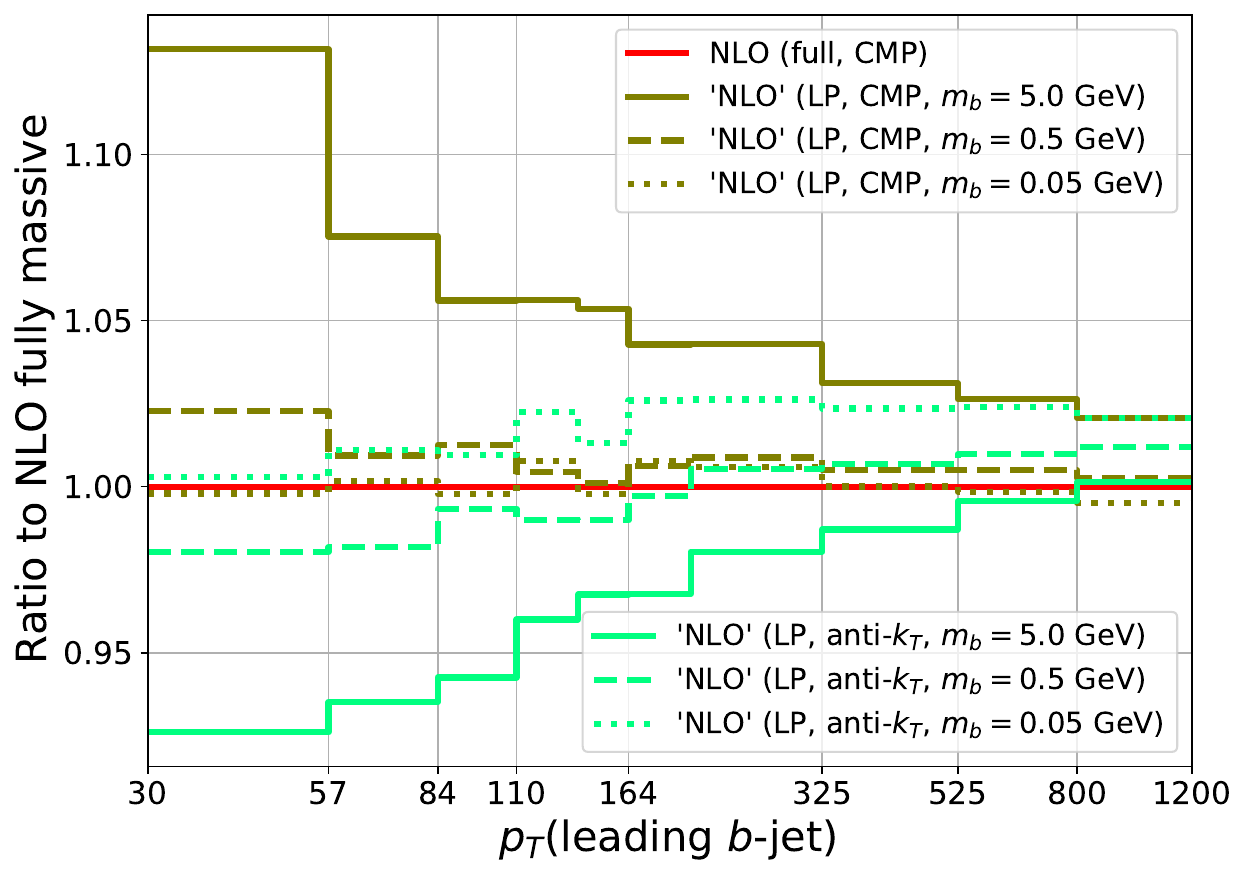}
	\caption{Left: ratios of the `NLO' LP anti-$k_T$ (green) and CMP (brown) cross sections to the NLO fully massive anti-$k_T$ cross section (red) for $m_b=5.0$ (solid), $0.5$ (dashed) and $0.05$ (dotted) GeV. Note that the mass is changed in tandem in both the numerator and the denominator. Right: the same as on the left, but as ratios to the NLO fully massive CMP cross section (red).}
	\label{fig:JFTppZbLP}
\end{figure}

The results are shown in the left panel of fig.~\ref{fig:JFTppZbLP}. Shown are `NLO' predictions for three values of the $b$-quark mass: $5.0$ GeV, $0.5$ GeV and $0.05$ GeV. All predictions are shown as ratios w.r.t.~the corresponding fully massive computations obtained using standard anti-$k_T$. The value of the mass used for the massive calculation in the denominator is always matched to the value used for the LP result in the numerator. For a mass of $5.0$ GeV, the results obtained using CMP differ from the LP results obtained using standard anti-$k_T$ by around 20\%, in line with the previously observed 10\% deviation. The fact that we do not get the exact same deviation is not surprising, since the relative weight of different contributions for $n_f=4$ and $n_f=5$ can be very different. What is interesting, however, is that the `correct' result, i.e.~the fully massive result, happens to lie almost exactly in the middle between the two LP predictions. This is of course a numerical coincidence, but it highlights that neither prediction is accurate at low $p_T$: the presence of power corrections on the order of $5-10\%$ limits the accuracy of any massless computation in this regime. This means that, to some extent, it is pointless to argue over which approach to defining jet flavour is best, if different approaches typically only differ by a few percent - an effect which is negligible compared to the size of power corrections which all approaches neglect. Furthermore, this also means that if a precision greater than about $5\%$ is desired, then this leaves us with no choice but to compute (at least part of) the cross section using massive quarks. This conclusion is consistent with the results of ref.~\cite{Gauld:2020deh}, which also found that the power corrections in this process are on the order of 5\%.

Considering now the curves corresponding to other values of the $b$-quark mass shown in the left panel of fig.~\ref{fig:JFTppZbLP}, we can see that, as the mass is reduced, the LP anti-$k_T$ cross section approaches the fully massive one, as expected. It is interesting to note that the rate at which the two curves approach each other, however, is extremely slow. One would naively expect the power corrections to already be completely negligible for a mass of $0.5$ GeV, even for the lowest transverse momenta shown. However, as it turns out, they are still clearly visible. It would appear - perhaps unsurprisingly - that this cross section suffers from linear power corrections, which at higher orders are further enhanced by logarithms of the $b$-quark mass. As a result, the power corrections only decrease by a factor of four or so as the mass is reduced from $5.0$ GeV to $0.5$ GeV.

Shifting focus to the CMP curves, we can see that they do not approach the fully massive result in the small-mass limit. Instead, each bin is shifted down at high $p_T$ by a roughly fixed amount for each factor of ten by which the mass is reduced. This is exactly the behaviour expected from results which differ by a single logarithm of the mass. We can also see that, at low $p_T$, the first factor of ten has a larger effect than the second. This suggests that the power corrections are negative for CMP, while they were positive for anti-$k_T$. This thus already explains a large portion of the 10\% difference between the two algorithms observed before.

The power corrections being negative for CMP can also be explicitly seen in the right panel of fig.~\ref{fig:JFTppZbLP}, which shows the same curves as the left panel, but this time as ratios to the fully massive CMP cross sections. As expected, the LP CMP curves properly converge to the fully massive result, while the LP anti-$k_T$ curves do not. This shows that for both anti-$k_T$ and CMP, the LP results are indeed LP-accurate. This constitutes a further test of the implementation, along the same lines as the lepton-collider test in the previous section, except now for an LHC process. The $m_b=0.05$ GeV results would not have been feasible without the use of the optimisation technique described in appendix \ref{sec:appendix}, which sped up the calculation by roughly two orders of magnitude.\footnote{Within this optimisation approach, the NLO fully massive calculation in this particular case is treated as an `NNLO' calculation, since its (quasi-)singularity structure is that of an NNLO cross section. This is the same reason why the `NLO' LP cross sections had to be computed using NNLO techniques, as mentioned before.}

\begin{figure}[t]
	\centering
	\includegraphics[width=0.49\textwidth]{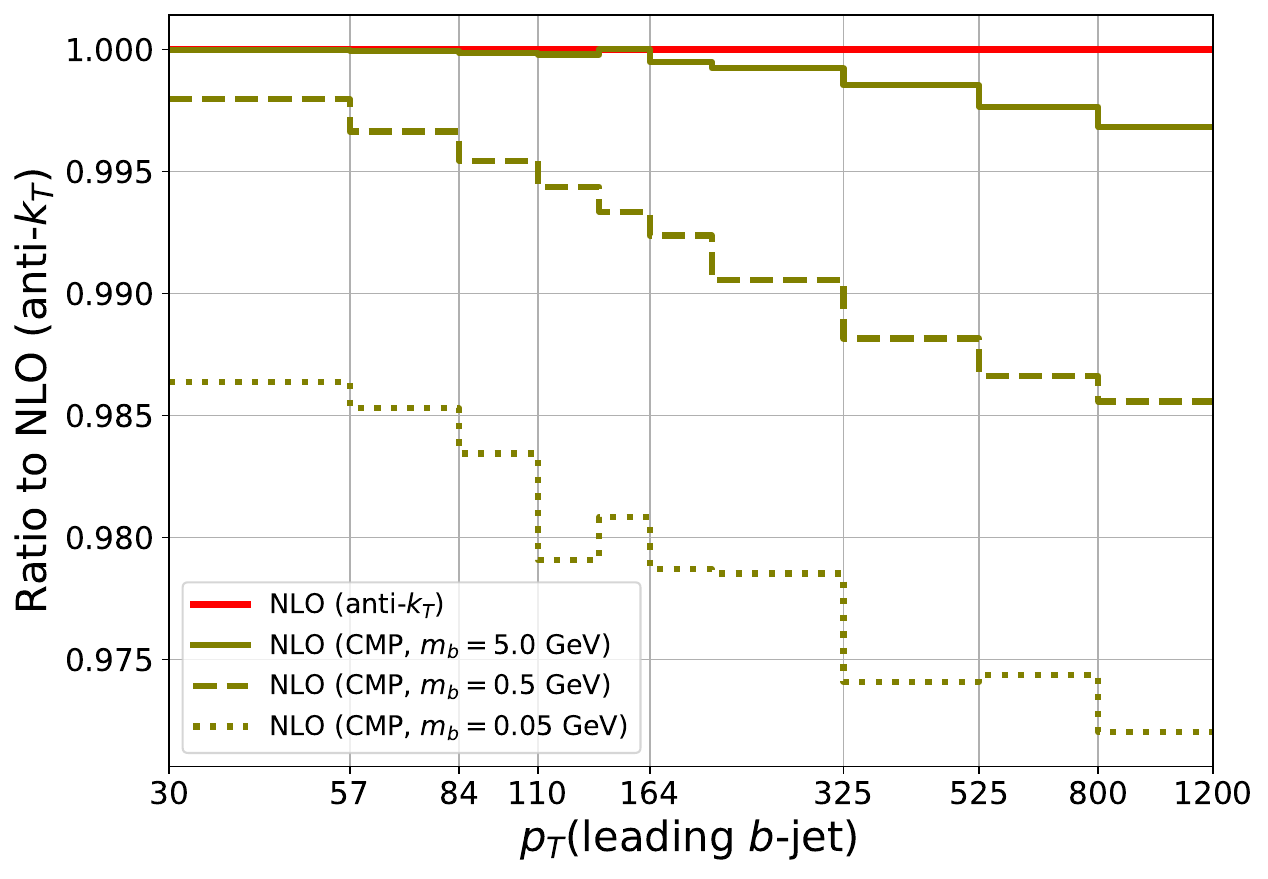}
	\caption{Comparison of the NLO fully massive cross sections obtained using anti-$k_T$ (red) and CMP (brown) for $m_b=5.0$ (solid), $0.5$ (dashed) and $0.05$ (dotted) GeV.}
	\label{fig:JFTppZbLP2}
\end{figure}

The use of anti-$k_T$ and CMP leading to opposite-sign power corrections in the mass at least partially explains the 10\% difference between the corresponding LP cross sections. However, this does not rule out the possibility that the effective energy cutoff of CMP for $a=0.1$ is too low at low $p_T$. To test this conclusively, we can study the difference between CMP and anti-$k_T$ jets for fully massive computations. If the effective energy cutoff introduced by CMP exceeds that of the mass, then switching from anti-$k_T$ to CMP reduces the cross section. If it is lower or the same, then the cross section remains unaffected by this change. Such a study is shown in fig.~\ref{fig:JFTppZbLP2}, which shows the ratio between fully massive results obtained using CMP and anti-$k_T$. For a mass of $5.0$ GeV, there is no difference between the two cross sections up until around $160$ GeV, at which point the CMP cross section starts to be suppressed relative to the anti-$k_T$ one. This suggests that around $160$ GeV, the effective energy cutoff on soft $b$-quarks introduced by CMP using $a=0.1$ matches the physical energy cutoff. For lower transverse momenta, a massless calculation would thus be allowed to probe an unphysical region of the phase space, yielding overestimated cross sections. This statement of course depends on the process, phase space cuts, etc., but it highlights that great care must be taken when employing such algorithms, since their effective cutoff might be significantly lower than initially expected. Note that these results are consistent with some of the results of ref.~\cite{Behring:2025ilo}: e.g.~fig.~16 of that work also shows that the impact of switching jet flavour definitions is much smaller for cross sections based on massive quarks than for those based on massless quarks, hinting at the same problem.

The curve for $m_b=0.5$ GeV suggests that for that mass, the point where CMP potentially starts probing flavoured quarks with unphysical kinematics lies around 40 GeV. This is roughly in line with our earlier observation that the effective mass introduced by CMP for $a=0.1$ is about $4\cdot10^{-3}\cdot40$ GeV $=0.16$ GeV for $p_T=40$ GeV. Correcting this estimate for the different power corrections of CMP and anti-$k_T$ has increased this to $0.5$ GeV, which is still very low.

Finally, looking at the curve for $m_b=0.05$ GeV, we can see that it is shifted relative to the curve for $m_b=0.5$ GeV by a constant. In fact, at the highest transverse momenta, the relative shift is the same every time the mass is reduced by a factor of ten. This is again as expected from cross sections which differ by a single logarithm of the quark mass. Since the curve for $m_b=5.0$ GeV smoothly transitions to unity for decreasing transverse momentum, a better estimate of where CMP with $a=0.1$ starts probing unphysical kinematics is obtained by finding where the $5.0$ GeV curve would intersect unity if it were obtained by adding the relative shift between the $0.05$ GeV and $0.5$ GeV curves to the $0.5$ GeV one. This point would lie around roughly $600$ GeV. This is again consistent with our earlier estimate of the effective mass introduced by CMP, up to a factor of two or so.

The opposite-sign power corrections and the large unphysical logarithms potentially introduced by CMP both explain why, contrary to the expectation, using flavoured jet algorithms can lead to higher cross sections than using anti-$k_T$, as observed in fig.~\ref{fig:JFTppZbRene}. It is worth noting that, contrary to our findings at LP NNLO, ref.~\cite{Behring:2025ilo} found that flavoured jet algorithms do indeed reduce the cross section relative to anti-$k_T$ in parton shower simulations. However, some of those simulations used massive quarks, while in others the relevant IR singularity was screened by a shower cutoff, lest the anti-$k_T$ cross section would not even be well-defined. Either way, the singularity was first screened by some other means before applying each jet algorithm. In that case, the $b$-quarks are treated - at least in some effective sense - as massive, and the use of flavoured jet algorithms must necessarily lead to a reduced cross section. Indeed, the fixed-order results shown in fig.~\ref{fig:JFTppZbLP2} for massive $b$-quarks also follow this expectation.

We now return briefly to the size of the mass logarithms. It is worth noting that, while fig.~\ref{fig:JFTppZbRene} suggests their contribution to be very small for all transverse momenta shown, this may very well cease to be true at higher orders. Indeed, at NNLO, the mass logarithm is essentially proportional to the LO cross section. However, this process receives very large NLO corrections at high transverse momenta, with a K-factor of around 10 at 1 TeV. Ignoring other types of logarithmic contributions that show up at higher orders, this would naively imply the logarithmic contribution to be 10 times larger at N$^3$LO for 1 TeV jets, reaching roughly 10\%. This would be consistent with the results of ref.~\cite{Behring:2025ilo}, which indicate that describing this process at high transverse momenta using parton showers leads to significantly larger differences of $\mathcal{O}(10\%)$ between algorithms than those observed for an NNLO calculation, which are typically at the percent level. Until an N$^3$LO computation is performed, one cannot be certain that the effect is actually that large, but if it is, then this would be sufficiently large to either require resummation or to demand the use of a flavoured jet algorithm to eliminate the logarithms entirely. However, we point out that, following the discussion above, some flavoured jet algorithms might suffer from the same problem. For example, if the interpretation that CMP does not sufficiently screen unphysical phase space regions is correct, then the corresponding NNLO logarithms of the effective quark mass would also be proportional to the LO cross section and be enhanced by the same factor of 10 at higher orders. Overall, while potentially much larger logarithmic effects would be unfortunate, they do not alter the above discussion qualitatively: if a given flavoured jet algorithm does not properly avoid probing unphysical kinematics, then it generally introduces larger logarithms than using anti-$k_T$ at LP, regardless of how large the mass logarithms of anti-$k_T$ may be. However, if a given algorithm can be shown not to suffer from this problem, then the absence of such large logarithms would certainly be a major advantage over using anti-$k_T$.

\subsection{Single-inclusive $b$-jet production at the LHC}

We will now consider the inclusive $b$-jet $p_T$ spectrum at the LHC, using the same setup already discussed in the previous section in the context of numerical checks. To our knowledge, this cross section has not been considered at NNLO before. As before, missing-higher-order uncertainties are estimated by performing standard 7-point scale variation. Since we will not be considering fully massive cross sections in this subsection, it is understood that all results correspond to LP cross sections.

\begin{figure}[t]
	\centering
	\includegraphics[width=0.49\textwidth]{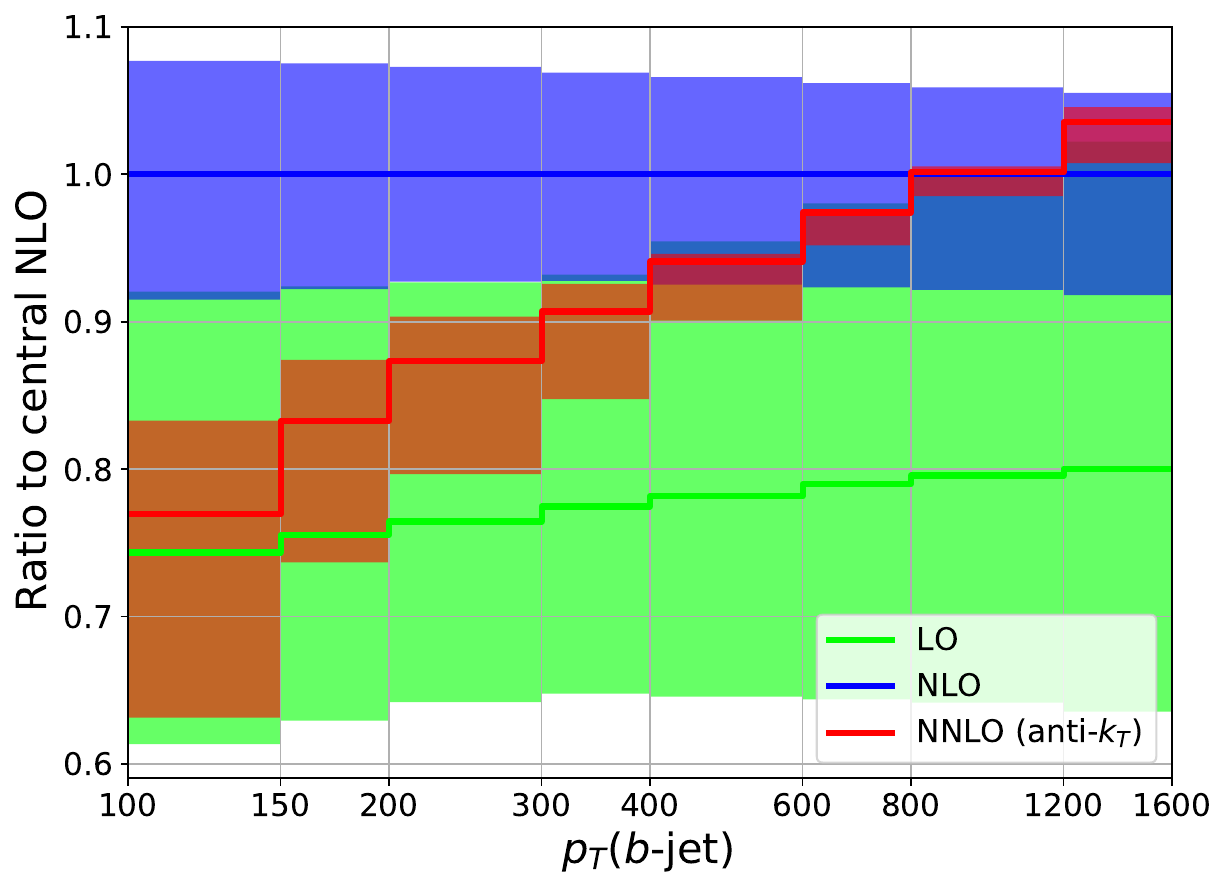}
	\includegraphics[width=0.49\textwidth]{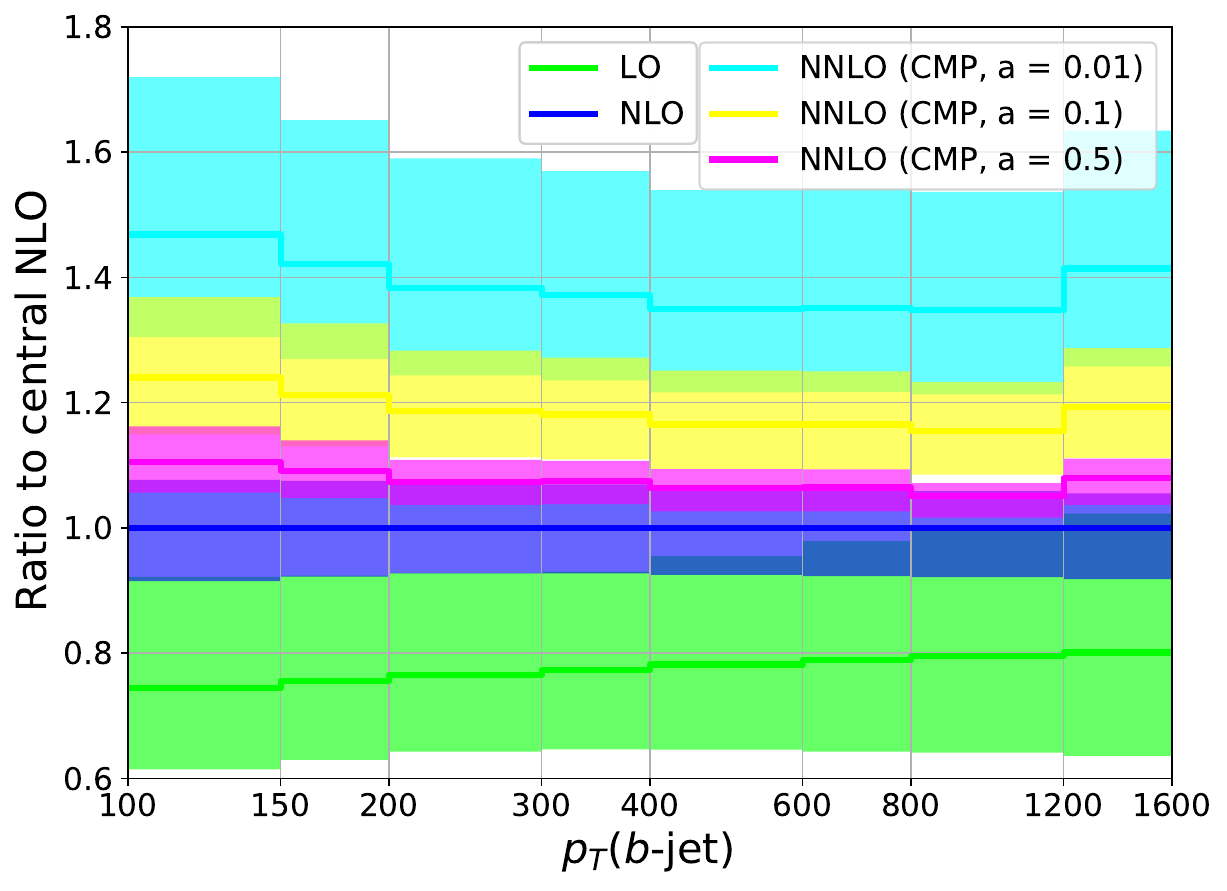}
	\caption{Left: the single-inclusive $b$-jet cross section using anti-$k_T$ at LO (green), NLO (blue) and NNLO (red). Error bands indicate the scale uncertainties. Right: as on the left, but at NNLO using CMP with $a=0.01$ (cyan), $a=0.1$ (yellow) and $a=0.5$ (purple).}
	\label{fig:JFTppb}
\end{figure}

The left panel of fig.~\ref{fig:JFTppb} shows the results obtained using anti-$k_T$ through NNLO. The NLO curve is only marginally consistent with the LO result and the NNLO curve shows very poor perturbative convergence below $300$ GeV. The latter is evident both from the scale uncertainties, which do not decrease after adding the NNLO corrections, and from the size of the NNLO corrections, which exceed the size of the NLO scale band. Additionally, the NNLO/NLO K-factor is significantly less flat than the NLO/LO K-factor. We confirmed that increasing the mass again decreases the NNLO cross section, worsening the perturbative convergence. The `large' logarithm of the $b$-quark mass is therefore again not to blame. Instead, it is known \cite{Currie:2016bfm,Currie:2018xkj,Bellm:2019yyh,Generet:2025vth} that single-inclusive jet cross sections exhibit poor perturbative convergence, or rather, that the scale uncertainties give an inaccurate impression of the size of missing higher orders for these processes. It would seem that our results confirm this to be the case for single-inclusive flavoured jet cross sections as well.

Interestingly, the situation looks different for the cross section obtained using the CMP algorithm. This is shown in the right panel of fig.~\ref{fig:JFTppb}. We consider cross sections for $a=0.01$, $0.1$ and $0.5$. Clearly, the cross section for $a=0.01$ converges very poorly: the NNLO scale uncertainties are much larger than at NLO and the NNLO curve lies well outside the NLO scale band. This is to be expected, since, at LP, the CMP cross section diverges in the $a\to0$ limit. The $S$ function counterterm which renders the anti-$k_T$ result finite is always zero for CMP for any $a>0$, since it only contributes in the strict double-soft limit. Meanwhile, for arbitrarily small non-zero values of $a$, arbitrarily soft quark pairs in the massless cross section can still affect the flavour of hard jets, effectively introducing a large log $\ln a$ with a coefficient roughly proportional to the LO cross section. Interestingly, the NNLO/NLO K-factor is very flat for all shown values of $a$. As the value of $a$ increases, the apparent convergence significantly improves. For the unusually large value $a=0.5$, the series naively appears to be convergent. However, given the long history of incorrect conclusions based on apparent perturbative convergence that has plagued this class of processes, we do not wish to jump to any conclusions based on such criteria alone.

Another interesting aspect is the strong dependence of the NNLO cross section on $a$. In ref.~\cite{Czakon:2022wam}, the effect on the $Z+b$-jet cross section of changing $a$ from $0.1$ to $0.5$ was found to be just a few percent at high $p_T$. For single-inclusive $b$-jet production, the effect is roughly 15\% - significantly more sizeable. This suggests this process is much more sensitive to the choice of jet flavour definition, making it a potentially interesting test bed for future comparisons of different approaches to jet flavour.

Comparing now the two algorithms, we reach similar conclusions as in the previous subsection. First, CMP tends to yield higher cross sections than anti-$k_T$. At low $p_T$, the difference is probably again mostly the result of power corrections. But at high $p_T$, this seems quite unlikely. Interestingly, we found that changing $m_b$ to $\mathcal{O}($few $10^{-3})\times p_T$ again roughly reproduces the CMP result for $a=0.1$, despite all the differences between this process and $p\:p\to Z\:b$-jet.

Second, one needs to choose $a$ to be unexpectedly large in order for the two algorithms to give the same result. In fact, the differences are much larger for single-inclusive $b$-jet production than they are for $Z+b$-jet production: even for $a=0.5$, the two algorithms roughly agree for only the highest $p_T$ bin. Since such a large value of $a$ surely sufficiently shields against soft $b$-quark pairs with unremarkable kinematic configurations, this suggests that maybe a certain unusual limit is the cause of these differences.  Comparing the two algorithms sector by sector, we find that the biggest differences by far occur in the process $g\:g\to g\:g\:b\:\overline{b}$ for triple-collinear sectors involving a gluon reference and two unresolved $b$-quarks. These sectors obviously contain the problematic soft-quark-pair limit, but the differences for other sectors containing that singularity are much milder. This suggests that the unphysical kinematics probed by CMP are related to configurations where a gluon and a $b$-quark pair are close to collinear. Further investigating the details of this phenomenon would require a dedicated study beyond the scope of this work. Whatever the cause of the differences, their large size further supports the idea that this process is highly sensitive to the definition of jet flavour.

As a concluding remark, it would be interesting to extend the results of ref.~\cite{Generet:2025vth} to the flavoured case to see if including small-jet-radius resummation also improves the perturbative behaviour for single-inclusive flavoured jet production. Now that it is possible to use standard anti-$k_T$ clustering, this should be significantly more manageable.

\section{Conclusion}\label{sec:Conclusion}

We have demonstrated how to compute quark mass effects in flavour-modulo-2 flavoured jet cross sections at leading power. Incorporating these leading power effects yields an IRC-safe cross section. By analysing all sources of mass dependence, we were able to reduce the required amount of theory effort to a minimum: only simple modifications to suitably-constructed subtraction schemes are needed and a single simple additional contribution has to be implemented and computed. The numerical integration of this extra contribution converges much more quickly than the rest of the calculation, meaning there is not even any extra computational cost. Meanwhile, since our approach is compatible with all commonly used jet clustering algorithms and yields - up to power corrections - the cross section one would obtain if one were to compute the cross section using the physical quark mass, there is no need for experimental collaborations to adapt to a new and involved definition of jets or jet flavour. They can simply keep using anti-$k_T$, or any other jet algorithm of their choice. An unfolding to flavour modulo-2 scheme jets still needs to be performed to directly compare to theory predictions, but such an unfolding should be simplified by the uncomplicated nature of the approach presented here.

Through arguments based on the logarithmic structure of the cross section and using explicit phenomenological studies, we have demonstrated that this approach does not lead to logarithms of the heavy quark mass sufficiently large to pose a threat to perturbative convergence for phenomenologically relevant energies. Instead, we have found that, at typical LHC energies, power corrections in the mass easily exceed the size of such logarithms. This implies that, if one is to achieve a precision sufficiently high to notice these `large' logarithms, then one needs to take into account the full mass dependence of the cross section, at which point there is no need to use special flavoured jet algorithms.

The leading-power results show some unexpected differences between anti-$k_T$ and flavoured jet algorithms like IFN and CMP. First of all, the observed differences between anti-$k_T$ and either IFN or CMP are $\mathcal{O}(10\%)$, much larger than the typical differences between flavoured algorithms found in ref.~\cite{Behring:2025ilo}. Second of all, the use of flavoured algorithms leads to higher cross sections at low transverse momenta than the use of anti-$k_T$, contrary to the physical expectation and the parton-shower-based results of ref.~\cite{Behring:2025ilo}. We have found evidence that suggests these surprising features are partially caused by power corrections in the quark mass. Furthermore, our results suggest some flavoured jet algorithms may not necessarily prevent unphysical regions of the phase space from being probed. This introduces large logarithms which exceed those related to the finite quark mass. Aside from obviously being an undesirable property, this is another potential explanation for why flavoured jet algorithms can lead to higher cross sections than anti-$k_T$ at leading power. Within the scope of this study, we have been unable to trace this potential pathological behaviour of some flavoured jet algorithms to specific kinematic regions. This issue certainly deserves a dedicated study testing multiple jet flavour definitions in this regard, which we leave for future work. Until a more complete understanding of the leading-power behaviour of jet algorithms - be it flavoured jet algorithms or anti-$k_T$ - at low transverse momentum has been obtained, we suggest any future leading-power results involving flavoured jets with transverse momenta up to several hundred GeV be cross-checked against a calculation retaining the full quark-mass-dependence. Alternatively, combinations of massive and massless calculations, like the one performed in ref.~\cite{Gauld:2020deh}, should be investigated.

We have also extended the sector-improved residue subtraction scheme and its implementation in \textsc{Stripper} to be capable of handling processes without massless partons at the Born level. Not only can this extension be used for phenomenological studies of such processes in the future, it can also be leveraged as an extremely powerful optimisation technique for cross sections containing quasi-singularities.

In the future, our approach can be used to obtain results for a wide range of jet processes at NNLO, which can be directly compared to measurements made using standard jet algorithms and the flavour modulo-2 scheme. We also expect our work to play an important role in ongoing discussions between theory and experiment on how to move forward regarding the jet flavour issue.

\begin{acknowledgments}

We are especially grateful to Micha\l{} Czakon for making the \textsc{Stripper} library available to us. We would also like to thank René Poncelet for a careful reading of the manuscript, for helpful discussions about literature results on jet flavour and for providing some of the numerical results of ref.~\cite{Behring:2025ilo}. We further thank Jingyu Zhang for a discussion on distinguishing quark and gluon jets. T.G.~has been supported by STFC consolidated HEP theory grants ST/T000694/1 and ST/X000664/1. This work was performed using the Cambridge Service for Data Driven Discovery (CSD3), part of which is operated by the University of Cambridge Research Computing on behalf of the STFC DiRAC HPC Facility (www.dirac.ac.uk). The DiRAC component of CSD3 was supported by STFC grants ST/P002307/1, ST/R002452/1 and ST/R00689X/1.

\end{acknowledgments}

\appendix

\section{Massive sectors}\label{sec:appendix}

One of the numerical checks presented in section \ref{sec:Checks} involved the computation of fully differential NNLO cross sections for the process $e^+e^-\to Q\overline{Q}$, where $Q$ is a massive quark. While such computations have been available for a while \cite{Gao:2014nva,Gao:2014eea,Chen:2016zbz}, they pose an interesting problem for the sector-improved residue subtraction scheme. Indeed, while the scheme and its implementation in \textsc{Stripper} are fully general, meaning they can be used to compute almost any cross section at NNLO, there is a very small set of processes they previously could not describe. This is because the construction of the subtraction scheme relies on defining sectors relative to massless reference partons. In very rare cases of phenomenological interest - essentially only $e^+e^-\to Q\overline{Q}$ - there are no massless partons at leading order. As a result, no sectors can be defined and the subtraction terms cannot be constructed according to the sector-improved residue subtraction scheme.

Thus, for the purpose of the present study, we extended the definition of the subtraction scheme and its implementation in \textsc{Stripper} to be able to handle Born processes without massless partons, covering this final edge case and finally rendering the subtraction scheme truly general - without exceptions. To do this, we use the following two-step procedure:
\begin{enumerate}
\item Pretend the massive partons are massless and construct the $n+2$ phase space accordingly, using the conventional phase space parametrisation of the subtraction scheme.
\item Massify the phase space using the RAMBO algorithm \cite{Kleiss:1985gy}.
\end{enumerate}
While seemingly incredibly simple, there are several technical details that require great care. This approach also has a major numerical advantage, which we will discuss later.

First, let us consider this approach at NLO. The premise is that there are only massive partons at LO, so at NLO there are only massive partons and one single gluon. Clearly, by first taking the massive partons to be massless, at least one sector can always be defined, allowing us to use the usual construction of the subtraction scheme to generate the massless phase space. The resolved $(n+1)$-particle phase space generated by the two-step procedure is trivially correct: we start from a correct and complete parametrisation of the massless phase space, so the RAMBO procedure is guaranteed to yield the correct corresponding massive one. This only leaves the (integrated) subtraction terms. Clearly, there are no (soft-)collinear singularities, so the corresponding (integrated) subtraction terms will have to be removed. There is a soft singularity, however due to the simple and fully numerical nature of the subtraction scheme - defining each subtraction term in a single strict limit and integrating over all other variables numerically - there is no need to redefine subtraction terms: they are defined point-by-point in the phase space using the particles' actual momenta in the limit, instead of relying on complicated momenta mappings. Thus, they do not care about how a phase space point was generated, only what it looks like in the relevant limit. Since the integrated subtraction terms involve no analytical integration to begin with, there is also no need to recompute any analytical integrals. The only point of care here is the strict limit: although the gluon is infinitely soft and does not contribute to observables, its momentum must also be rescaled during the RAMBO procedure, just like all the non-vanishing momenta. This affects both the phase space weight of the subtraction term (through the RAMBO reweighting factor) and the factorised soft singularity, since it depends on the regulated gluon energy ($u^0/\xi$ in the notation of ref.~\cite{Czakon:2014oma}). Similarly, the gluon should be treated as $d$-dimensional for the integrated subtraction term. This only affects the phase space weight through the substitution\footnote{Here `$\xi$' is the factor by which 3-momenta are rescaled during the RAMBO massification, not the variable mentioned a few lines earlier parametrising the unresolved parton's energy within \textsc{Stripper}.}
\begin{equation}
\xi^{2n-3} \to \xi^{2n-3-2\varepsilon}
\end{equation}
in the reweighting factor of the RAMBO procedure.

At NNLO, two things change. First, there are now up to two soft partons at a time. However, there are no extra modifications necessary beyond combining those for each individual soft parton. The second difference is the presence of a collinear singularity between the two additional massless partons. However, as before, the implementation of the corresponding subtraction terms in $\textsc{Stripper}$ is done directly based on the generated momenta, not on the initial phase space parameters, so once again no further modifications are necessary. One should just take care to reweight the phase space using the RAMBO formula for $n+2$ partons, rather than treating the two collinear partons as a single parton.

This strategy is thus incredibly simple and elegant. Perhaps its greatest feature, however, is not that it finally enables NNLO computations of processes with no massless partons at the Born level using the sector-improved residue subtraction scheme. Instead, its greatest strength is its application as a powerful optimisation technique for cross section calculations involving very light massive quarks. For example: for $e^+e^-\to Q\overline{Q}$, it introduces two sectors at NLO: one sector parametrising the gluon relative to the quark and one sector parametrising the gluon relative to the anti-quark. While there is no collinear singularity between the massive quarks and the gluon, there is a quasi-collinear singularity if the mass of the quark is small compared to its energy. While there is no need to define the gluon's direction relative to anything in this process - one could just sample its direction uniformly and only worry about parametrising its energy in a way that facilitates the use of a subtraction scheme - it is incredibly useful to do so. If the gluon's direction were sampled uniformly, then these quasi-collinear singularities would be located on complicated hyperplanes within the $(0,1)^m$ hypercube of integration variables. From a Monte Carlo perspective, if the mass is small enough, this would appear as very strongly peaked structures within the hypercube, which are only sampled rarely and randomly. This leads to very poor numerical convergence of the Monte Carlo integration.

The construction of the phase space within the sector-improved residue subtraction scheme directly parametrises the angle between the reference and unresolved partons in a sector. Additionally, the selector functions guarantee that a given quasi-collinear singularity is contained within a single sector. A given quasi-collinear singularity is thus located entirely on one of the faces of the hypercube of a single sector, making it trivial to importance sample the peaked behaviour of the cross section. This is especially impactful at NNLO, where the quasi-singularity structure is much more complicated, with nested quasi-collinear limits and double-quasi-soft massive quark pairs.

Using this massification procedure makes importance sampling all these quasi-singularities trivial, speeding up the Monte Carlo convergence by several orders of magnitude - without any extra effort from the user - if the mass is small enough. This is often the case for numerical checks, like those performed in this paper, where minimising the Monte Carlo errors is very important, but typically challenging. However, even for phenomenological studies of e.g.~massive $b$- or $c$-quark production at high energies, this technique can easily yield significant speed-ups.

This now suggests applications of the massification procedure to processes at hadron colliders, even though this is strictly speaking not necessary, since in those cases there will always be massless partons at the Born level. Indeed, this was used in section \ref{sec:Pheno} to compute NLO cross sections for $p\:p\to Z\:b\:\overline{b}$ with very light $b$-quarks. In principle, this is straightforward: one simply adds the extra sectors involving massive partons, treating these massive sectors as described above. The original sectors without any massive reference or unresolved partons can be treated as usual, without the need to massify the phase space using a two-step procedure.\footnote{Although one certainly could do so, if desired.}

However, applying the procedure to hadron colliders comes with a few more minor complications. First, the centre-of-mass energy of the massless phase space point might be below the threshold of the massive phase space. One might be tempted to simply sample the PDF fractions in a way that guarantees that the total centre-of-mass energy of the initial massless phase space is sufficient. However, this is not that trivial in practice. If the sector involves final-state references only, then this will work trivially. However, if one of the reference partons is in the initial state, then, within the sector-improved residue subtraction scheme, the cross section and its subtraction terms will have different centre-of-mass energies, complicating the minimum-energy requirement. While it is not terribly difficult to simply come up with alternate phase space parametrisations for these cases, we instead opted to simply sample all centre-of-mass energies and reject contributions with insufficient energy. While technically less efficient - sometimes generating phase space points that cannot contribute to the cross section - in practice this inefficiency is negligible. Indeed, the premise for using this technique is that the massive quarks are light relative to their energy. This makes the difference between the initial massless phase space point and its massified version tiny. Thus, configurations with insufficient energy will not or only rarely be sampled anyway.

The final complication caused by initial-state references is due to certain assumptions usually built into subtraction scheme codes. Typically, if a parton is taken to be collinear to the initial state, it is essentially removed from the final state - it cannot be observed anyway. However, for the purpose of the massification, it has to be kept, just like for final-state collinear singularities. Furthermore, since a gluon becoming soft or collinear to an initial-state reference usually yields the same observable phase space point within the sector-improved residue subtraction scheme, there is typically no need to recompute the observable momenta for each of those limits separately. However, when using the massification procedure, they do differ: the energy of the gluon affects the massification of the entire phase space. Thus, there is an observable difference between the configurations with a non-soft collinear gluon and a soft gluon. This is not a problem for a final-state collinear singularity, since in that case the sum of the energies of the reference and unresolved partons does not depend on the softness of the collinear gluon, and so the observable phase space is unaffected.

\end{document}